\documentclass[a4paper,11pt]{article}

\usepackage{jheppub}

\usepackage[T1]{fontenc}

\usepackage{graphicx}

\title{Quantum-induced interactions in the moduli space of degenerate BPS domain walls}

\author[a]{A. Alonso-Izquierdo,\note{Corresponding author.}}
\author[b]{J. Mateos Guilarte,}

\affiliation[a]{Departamento de Matematica Aplicada and IUFFyM \\ Universidad de Salamanca, SPAIN}
\affiliation[b]{Departamento de Fisica Fundamental and IUFFyM \\ Universidad de Salamanca, SPAIN}

\emailAdd{alonsoiz@usal.es}
\emailAdd{guilarte@usal.es}

\abstract{In this paper quantum effects are investigated in a very special two-scalar field model having a moduli space of BPS topological defects. In a $(1+1)$-dimensional space-time the defects are classically degenerate in mass kinks, but in $(3+1)$ dimensions the kinks become BPS domain walls, all of them sharing the same surface tension at the classical level. The heat kernel/zeta function regularization method will be used to control the divergences induced by the quantum kink and domain wall fluctuations.  A generalization of the Gilkey-DeWitt-Avramidi heat kernel expansion will be developed in order to accommodate the infrared divergences due to zero modes in the spectra of the second-order kink and  domain wall fluctuation operators, which are respectively $N\times N$ matrix ordinary or partial differential operators. Use of these tools in the spectral zeta function associated with the Hessian operators paves the way to obtain general formulas for the one-loop kink mass and domain wall tension shifts in any $(1+1)$- or $(3+1)$-dimensional $N$-component scalar field theory model. Application of these formulae to the BPS kinks or domain walls of the $N=2$ model mentioned above reveals the breaking of the classical mass or surface tension degeneracy at the quantum level. Because the main parameter distinguishing each member in the BPS kink or domain wall moduli space is essentially the distance between the centers of two basic kinks or walls, the breaking of the degeneracy amounts to the surge in quantum-induced forces between the
two constituent topological defects. The differences in surface tension induced by one-loop fluctuations of BPS walls give rise mainly to attractive forces between the constituent walls except if the two basic walls are very far apart. Repulsive forces between two close walls only arise if the coupling is approaches the critical value from below.}


\begin{document}

\maketitle
\flushbottom


\section{Introduction}

Domain walls are topological defects owing  their existence to the spontaneous symmetry breaking of a discrete group. These two-brane objects arise in a minimal scenario in one-real scalar field theory and have important implications in
areas as diverse as Cosmology and Condensed Matter Physics, see e.g. Reference \cite{Vilenkin}. In Reference \cite{Voloshin} Shifman and Voloshin discovered that topological objects of this type exist forming families of infinite BPS walls, degenerate in surface tension, in a ${\cal N}=1$ supersymmetric Wess-Zumino model with two chiral superfields, whereas Eto and Sakai showed in \cite{Eto} that families of degenerate domain walls also arise as exact solutions in ${\cal N}=1$ supergravity. In a parallel development, the same domain wall solutions were considered in a
purely bosonic context and in $(1+1)$-dimensional space time in the disguise of kinks. First, in \cite{Bazeia1997} two kinds of topological kinks were unveiled, either having only one non-null component of the iso-spin doublet scalar field or living on a half-elliptical orbit in field space. Second, in the paper \cite{Alonso2002} all the BPS kink orbits -henceforth, all the BPS domain wall orbits- were identified and shown to be identical to the topological wall orbits found in \cite{Voloshin}. Moreover, in \cite{Alonso2002} analytical expressions for the domain wall profiles, not only the orbits in field space, were
obtained for two critical values of the coupling between the two scalar fields. At these critical values, the
mechanical system of two degrees of freedom equivalent to the search for static topological walls is completely integrable.

All these BPS topological defects, either kink or domain walls, fluctuate along flat directions of the potential energy in the configuration space, i.e., they support zero modes. In fact, given one degenerate BPS topological defect there are two linearly independent zero modes: the translational mode, a null energy fluctuation due to the free motion of the extended solution center, and a Jacobi field due to the freedom of moving inside the moduli space from solution to solution. It was proposed by Manton, see \cite{Manton1982,Manton2004} that the adiabatic motion of BPS solitons can be modeled as geodesic motion in the moduli space equipped with a metric induced by the zero modes. Manton's approach was implemented in \cite{Alonso2002b} in the two-scalar field model in order to describe the low energy dynamics of these BPS kink defects. One of the zero modes responds to the free dynamics of the center of mass of the constituent lumps. The second zero mode is due to the
motion in the relative coordinate and induces a non-Euclidean metric in the moduli space parametrized by this relative coordinate between the two basic lumps. In this way, Manton's method unveils the low-energy one-dimensional scattering
of the elementary or constituent kinks, and, by promoting the whole construction to $(3+1)$ dimensions, the domain wall adiabatic motion in the transverse direction. In Reference \cite{Tong2002} Tong developed a similar analysis on the richer
moduli space of BPS walls arising in ${\cal N}=1$ supersymmetric quantum electrodynamics, whereas in \cite{Hindmarsh2003} Hindmarsh et al. studied the low-energy dynamics of kinks as a model for three-branes in $M$-theory.

The main theme in this paper is to investigate how the above described scenario is modified by quantum effects. Of course, zero modes give rise to quantum fluctuations. In this sense, Manton's geodesic dynamics is a \lq\lq pre-quantum\rq\rq effect. Our goal, however, is to take into account
alternatively higher-energy kink or domain wall fluctuations up to one-loop order. Regarding the $(1+1)$-dimensional context, the procedure established by Dashen, Hasslacher and Neveu in \cite{Dashen1974} to compute the one-loop kink mass shifts by developing the $\hbar$-expansion around the extended classical solutions in the $\phi^4$ and sine-Gordon scalar field models sets the standards of the topic. The DNH formula encodes the shifts in the classical kink energies induced by one-loop fluctuations by collecting three contributions : 1) the kink zero-point energy, the energy of the kink ground state where all the fluctuation modes are unoccupied, 2) the vacuum zero-point energy that must be subtracted from the kink zero-point energy, and 3) the energy induced by the one-loop mass renormalization counter-term on the kink background (measured with respect to the same effect on the vacuum). Even though the issue of quantum corrections to kink masses was placed on firm grounds, mainly by Dashen, Hasslacher and Neveu, in the seventies, a revival in the subject took place around the change of century. The interest
in computing the one-loop mass shifts for supersymmetric kinks again pushed forward the topic in supersymmetric theories \cite{Goldhaber2004, Shifman1999, Graham1999}. The delicate balance between the chosen regularization procedure before subtracting the zero-point vacuum energy and supersymmetry breaking required a careful rethinking of the DHN formula within the purely bosonic framework. It was found, see references \cite{Rebhan1997,Nastase1999}, that the regularization implicit in the DHN formula could be achieved by setting a cut-off in the number of fluctuation modes accounted for -rather than in the energy- and the result obtained in this way agrees with the exact result obtained in the completely integrable sine-Gordon model for the sine-Gordon kink.

We shall concentrate in the computation of the shifts in the surface tension of the degenerate domain walls of the model discussed in \cite{Alonso2004b}, the bosonic sector of the Shifman-Voloshin model \cite{Voloshin}. A natural question emerges: Is the classical domain wall surface tension degeneracy broken at the quantum level? We found hints in \cite{Alonso2004b}, see also \cite{Guilarte2010} to find a comprehensive review, that the answer is affirmative for classical kink masses but lack of control of the zero mode fluctuations at that time, a weakness of our method that we shall try to amend in this paper, prevented us from claiming a clear-cut result. In fact, the surge of quantum vacuum forces between topological solitons \cite{Bordag2012} or compact objects \cite{Castaneda2013} is a central issue in quantum field theory under the influence of external conditions and is the problem that we shall address regarding the two constituent lumps of our composite, first, kinks, and, then, domain walls.

There is a relevant, almost insurmountable, difficulty in the application of the DHN formula to the BPS-topological defects in our model,  except for the simplest one, where the Hessian operator is a $2\times 2$-order diagonal matrix differential operator. There is insufficient spectral information about the rest of the non-diagonal matrix differential operators governing the fluctuations around the generic topological defect to apply the DHN formula effectively. We recall that the kink fluctuation operators around the $\phi^4$ and sine-Gordon kinks are ordinary Schr\"odinger operators of the P\"oschl-Teller type. The spectral problem of operators in this class, also arising in the SV-model as the diagonal components of the simple kink Hessian, is exactly solvable and thus the DHN formula is fully applicable. The only alternative way to deal with this problem when the details about eigenvalues and eigenfunctions are unknown is to rely on the spectral functions such as the heat trace and the spectral zeta functions, see \cite{Elizalde1994, Kirsten2002, Vassilevich2003}. The virtue of the heat trace is that it can be obtained directly from the potential, its derivatives, and products and powers of these quantities from the heat kernel high-temperature expansion, which is an asymptotic series in a (fictitious) inverse temperature, see \cite{DeWitt1965, Gilkey1984, Roe1988, Avramidi}. In reference \cite{Avramidi1995} the Gilkey-DeWit heat kernel expansion has been generalized to matrix differential operators. Therefore, one does not need to know the eigenvalues to find the spectral zeta function via Mellin's transform of the heat trace. Considering the spectral zeta function as the main tool in the approach to computing one-loop effects, one is almost forced to use the zeta function regularization procedure as the most appropriate method of control of the ultraviolet divergences, see \cite{Dowker1976} and \cite{Hawking1977}. This elegant procedure was used in the calculation of one-loop mass shifts for supersymmetric kinks in \cite{Bordag2002} and, in a purely bosonic context, helped us to achieve interesting results about kink mass shifts in models with only one scalar field in \cite{Alonso2002c} even though the DHN formula did not work. It is worth mentioning that not only the zero point kink and vacuum energies are regularized by going to a regular point in the complex $s$-plane of the corresponding spectral zeta function, but also the ultraviolet divergence appearing in the one-loop mass renormalization counter-term is regularized in the same way using the vacuum spectral zeta function. The physical value of $s$, the point in the $s$-complex plane where the divergent physical quantities are defined, is a pole of the spectral zeta functions involved but the remainders are such that the renormalizations performed prompt finite and correct results that can be checked in cases where the shifts are known by other methods, see e.g. \cite{Alonso2013}. Similar techniques were developed in \cite{Alonso2002d} to work the one-loop kink mass shifts in a model with two scalar fields but without degeneracy between the classical kink masses.

If the algebraic kernel of the differential operator is non-null, i.e., if there are zero modes in the spectrum, the
exact heat trace and the Gilkey-de Witt-Avramidi heat trace expansion differ at low temperatures, where the zero modes become dominant. Therefore, one must restrict the integration domain of the Mellin transform to a finite interval where the exact and asymptotic heat traces fit well. The poles of the spectral zeta function are captured in the high-temperature domain, i.e., it suffices to limit the integration interval in Mellin's transform of the heat trace to $[0,1]$ to find, e.g., anomalies induced by fluctuations in the ultraviolet spectrum. By doing this one neglects a portion of the entire part of the spectral zeta function, a bad option when one is dealing with zero-mode fluctuations of extended objects. We improved on the error admitted in this procedure in \cite{Alonso2011}, where an optimum choice of the integration domain in Mellin's transform of the heat trace is generated by means of a numerical algorithm. Any truncation at non-null low temperatures is not theoretically satisfactory. The standard Gilkey-de Witt expansion works fine in the whole temperature range only for operators with a strictly positive spectrum. In two recent papers \cite{Alonso2012,Alonso2013b} we proposed a modification of the Gilkey-DeWitt expansion to be adapted to operators having zero modes in their spectra. The new asymptotic expansion was worked on one scalar field kinks. The modified procedure is not only conceptually more satisfactory but also enhances the numerical precision in the computation of kink mass quantum corrections to a remarkable extent.

In Reference \cite{Alonso2004b} we relied on the standard heat kernel expansion to evaluate the one-loop kink mass shifts, neglecting the zero modes. The lack of precision in the data due to the truncation of the
temperature range in Mellin's transform frustrates a reliable conclusion about whether or not the classical kink energy degeneracy is preserved at the quantum level. Here, we shall first generalize to field models with two scalar fields the modification of the heat kernel expansion that accounts for zero modes and allows us to extend the Mellin transform to the whole temperature range safely. We shall then use the modified heat trace expansion to estimate the one-loop shifts in the kink masses. The outcome is remarkable: there exists a critical value of the coupling constant between the two scalar fields that separates two different phases. If the coupling constant is lower than this critical value then the two constituent lumps repel each other; otherwise, the two basic kinks attract mutually if they are close enough and repel each other if they are distant enough. At the critical value of the coupling constant the classical degeneracy in the kink mass is preserved. This picture resembles a very peculiar phase transition induced by quantum, rather than thermal, fluctuations.

After calculation of the one-loop BPS kink mass shifts in $\mathbb{R}^{1,1}$-Minkowski space-time we shall confront the computation of the quantum corrections to the BPS domain wall surface tension up to the semi-classical level. Domain wall fluctuations have been discussed, e.g., in \cite{Carvalho1986} and \cite{Parnachev2000}, although a comprehensive analysis of this subject has been achieved in \cite{Rebhan2002}. Use of dimensional regularization allowed the authors of this paper to determine respectively the one-loop shifts to the classical kink mass, the domain ribbon
length tension, and the domain wall surface tension in the $\lambda(\phi)^4_2$, $\lambda(\phi)^4_3$, and $\lambda(\phi)^4_4$ scalar field model in various dimensions. More recently, in \cite{Rebhan2009}, similar results has been obtained for the fundamental topological defect of the sine-Gordon, $\phi^4$, and $\mathbb{C}\mathbb{P}^1$ models, in its different forms depending on the dimension, both in purely bosonic and supersymmetric settings, at zero and finite temperature. Dimensional regularization is well suited to jump over physical dimensions and analyze one-loop shifts to classical topological bounds characterizing ${\rm p}$-branes of different ${\rm p}=d-1$ in a unified way. Zeta function regularization is a close cousin of dimensional regularization, also well suited for computing almost simultaneously quantum shifts to classical extended objects of different dimensions related through dimensional reduction. We shall accordingly use heat kernel/zeta function methods in the computation of one-loop shifts to the BPS topological defects understood as domain walls in $(3+1)$-dimensions. The domain wall fluctuations are governed in this case by matrix partial differential operators but because the background depends only on one coordinate the heat kernel and zeta functions can be easily worked out from the corresponding spectral functions of the kink fluctuation operators. A subtle point is that the contribution of the fluctuations  parallel to the wall is not fully compensated by the vacuum fluctuations due to the phase shifts induced after crossing the domain wall in the transverse direction.
The results on one-loop shifts to wall tensions are qualitatively similar to those on kink mass shifts. There are different shifts for different members of the wall family although they are weaker than kink mass shifts but, contrarily to the kink mass shifts,  the behaviour above and below the critical coupling $\sigma=2$ is similar for wall tension shifts.

We shall pursue this investigation as a necessary intermediate development before of embarking ourselves in the quantum treatment of the domain walls existing in the Ginzburg-Landau non-linear $S^2$-sigma model of Reference \cite{Alonso2010}. The structures of the vacuum orbits and the moduli space of degenerate BPS-domain walls in both
models are similar. The so-called tropical domain walls in \cite{Alonso2010} form a degenerate family of BPS-domain walls with similar properties to those exhibited by the topological walls to be discussed in this paper. One can safely establish their stability in both models by application of the Morse index theorem, see \cite{Alonso2008,Alonso2009}. In the non-linear sigma model, however, the analysis of domain wall fluctuations is more difficult because of the non-flat curvature in field space.

The organization of the paper is as follows: in Section \S.2 we first describe the general setting in the search for domain walls in $N$-component scalar field theories allowing for BPS bounds and equations. We then introduce the particular model that we are going to discuss: the bosonic sector of the Shifman-Voloshin model \cite{Voloshin}. This subsection will be followed by a rapid description of the moduli space of tension-degenerated BPS-domain walls as well as the presentation of the framework to analyze the small domain wall fluctuations. Section
\S.3 contains the main theoretical novelties in the paper: the Gilkey-de Witt heat kernel expansion is adapted to $2\times 2$ matrix differential operators whose spectra involve zero modes. The new heat trace expansion is Mellin's transform integrated over the whole temperature range to obtain the spectral zeta function. The usual zeta function regularization/renormalization procedures are then implemented to estimate the one-loop kink mass shifts by means of a truncated asymptotic series in the coefficients of the heat trace expansion. In Section \S.4, the new formula is applied to the evaluation of the one-loop kink mass correction where the DHN formula is not applicable. In Section \S.5 the previously developed
machinery is generalized to evaluate the one-loop surface tension shifts of the BPS domain walls. The sub-Section \S.5.1 offers the exact calculation of the tension semi-classical correction to the simplest BPS wall whereas sub-Section \S.5.2 is devoted to the application of the heat
kernel expansion to compute the wall tension shifts for several values of the coupling and distances between the basic walls.
Finally, in Section \S.6 we offer some conclusions and propose several prospects.

\section{Degenerate classical BPS-domain walls in a two-scalar field theory model}

\subsection{General field theoretical background and conventions}

The action governing the dynamics in a $(1+d)$-dimensional relativistic field theoretical model of $N$-scalar fields is of the form:
\begin{equation}
\widetilde{S}[\Psi]=\int \!\! \dots \!\! \int \, dy^0dy^1 \dots  dy^d\, \Big(\frac{1}{2}\sum_{a=1}^N\frac{\partial\psi_a}{\partial y_\mu}  \frac{\partial\psi_a}{\partial y^\mu}- \widetilde{U}[\psi_a(y^\mu)] \Big) \label{action50}
\end{equation}
with $a=1,\dots,N$ and $\mu=0,\dots ,d$. Here, $\Psi(y^\mu)=\left(\begin{array}{c} \psi_1(y^\mu)\\ \vdots \\
\psi_N(y^\mu)\end{array}\right): \mathbb{R}^{1,d} \, \longrightarrow \, \mathbb{R}^N$ is a $N$-component real scalar field while $y^0,\dots, y^d$ are local coordinates in the Minkowski space-time ${\mathbb R}^{1,d}$ equipped with a metric tensor $g_{\mu\nu}={\rm diag}(1,-1,\dots,-1)$, $\mu,\nu=0,1,\dots,d$. The Einstein convention is only used on the space-time variables. We shall work in a system of units where the speed of light is set at one, $c=1$, but we shall keep the Planck constant $\hbar$ explicit because we plan to investigate the one-loop corrections, proportional to $\hbar$, to the classical mass of kinks and the tension of the domain walls induced by quantum fluctuations. In this system of units, the physical dimensions of fields and parameters are: $[\hbar]=[\widetilde{S}]=M L$, $[y_\mu]=L$, $[\psi_a]=M^\frac{1}{2} L^{1-\frac{d}{2}}$, $[\widetilde{U}]=ML^{-d}$. The specific model that we shall address is characterized by two parameters, $m$ and $\lambda$, respectively carrying the following physical dimensions: $[m]=L^{-1}$, $[\lambda]=M^{-1}L^{-d}$. We define the non-dimensional coordinates, fields and potential energy density in terms of these parameters: $x_\mu=m \, y_\mu$, $\Phi= \frac{\sqrt{\lambda}}{m^{d-1}} \Psi$ and $U(\Phi)=\frac{\lambda}{m^{2d}} \widetilde{U}(\Psi)$. The static part of the energy is also proportional to the dimensionless energy functionals:
\begin{equation}
\widetilde{E}[\Psi]=\frac{m^d}{\lambda}E[\Phi]= \frac{m^d}{\lambda}\int\!\dots \! \int\! dx^1\dots dx^d \, \left[\frac{1}{2} \sum_{a=1}^N \vec{\nabla} \phi_a \cdot \vec{\nabla}\phi_a +U[\phi_a(x)] \right] \, \, \, ,  \label{energy}
\end{equation}
where $\vec{\nabla}f(x^1,\dots,x^d)=\frac{\partial f}{\partial x^1}\vec{e}_1+\dots+
\frac{\partial f}{\partial x^d}\vec{e}_d$ is the gradient of a function in $\mathbb{R}^d$ and $\vec{e}_j$, $j=1,\dots,d,$ is an orthonormal basis of vectors. The configuration space ${\cal C}$ of the system is in turn defined as the set of finite-energy field configurations at a fixed time $t=t_0$: ${\cal C}=\{\phi_a(t_0,\vec{x})\in {\rm Maps}(\mathbb{R}^{d},\mathbb{R}^N)/ E[\Phi]<+\infty\}$.

If the action (\ref{action50}) arises in the bosonic sector of a supersymmetric model of Wess-Zumino type, the energy density function $U(\Phi)$ factorizes in the form:
\begin{equation}
U(\Phi)=\frac{1}{2} \sum_{a=1}^N \frac{\partial W}{\partial \phi_a}\cdot \frac{\partial W}{\partial \phi_a} \hspace{0.4cm}.
\label{susy01}
\end{equation}
The function $W(\Phi): \mathbb{R}^N \rightarrow \mathbb{R}$ is usually referred to as the superpotential in the framework of supersymmetric field theory. The critical points of the superpotential, $\frac{\partial W}{\partial\phi_a}(\Phi^c)=0$, are the static and homogeneous solutions of the system. Subsequently, the set of absolute minima of $U(\Phi)$, ${\cal M}=\{\Phi^{c(i)}\,\,/ \,\,U(\Phi^{c(i)})=0\}$, engenders the set of degenerate vacua in the quantum version of the system. Assuming that ${\cal M}$ is a discrete set for later purposes, the small (quadratic) fluctuations $\delta \phi_{a\vec{k}} (x^0,\vec{x}) = e^{i\nu(\vert\vec{k}\vert)x^0} \xi_{a\vec{k}}(\vec{x})$ around any of these constant solutions are determined by the eigenfunctions
$\Xi_{\vec{k}}(\vec{x})=\left( \xi_{1\vec{k}}(\vec{x}) \cdots  \xi_{N\vec{k}}(\vec{x}) \right)^t$, $\vec{k}\in{\rm Vec}(\mathbb{R}^d)$
of the second-order vacuum fluctuation differential matrix operator
\[
\mathbb{L}_0= -\nabla^2 \mathbf{I}_{N\times N} + \mathbf{v}^2 \hspace{0.5cm} \mbox{where} \hspace{0.5cm}  \mathbf{v}^2={\rm diag} \{v_1^2,\dots,v_N^2\}  \hspace{0.5cm} \mbox{and} \hspace{0.5cm} v_a^2 = \frac{\partial^2 U}{\partial\phi_a^2}[\Phi^{c(i)}]\hspace{0.4cm}  .
\]
From the spectral relation $\mathbb{L}_0 \Xi_{\vec{k}}(\vec{x}) =\nu^2(\vert\vec{k}\vert) \Xi_{\vec{k}}(\vec{x})$, $N$ decoupled one-dimensional spectral problems arise, one for each component $\xi_{a\vec{k}}(\vec{x})$: $[\mathbb{L}_{0}]^a_a \xi_{a\vec{k}}(\vec{x}) = (-\nabla^2 + v_a^2) \xi_{a\vec{k}}(\vec{x}) =\nu_a^2(\vert\vec{k}\vert) \xi_{a\vec{k}}(\vec{x})$. The $\xi_{a\vec{k}}(\vec{x})=e^{i\vec{k}\cdot\vec{x}}u_a$ functions solve the one-dimensional eigenvalue problems provided that the dispersion relations $\nu_a^2(\vert\vec{k}\vert)=\vert\vec{k}\vert^2+v_a^2$ hold. In quantum theory, these fluctuation normal modes become the fundamental quanta of the system, $v_a$ giving the meson masses.

The next step is to investigate the presence of topological defect solutions. In particular, we shall focus our attention on domain wall defects. Domain walls are smooth solutions of the field equations such that their energy density is a localized function in the $x^1$ direction and has a space-time dependence of the form ${\cal E}(x^0,x^1,x^2,\dots, x^d)={\cal E}(x^1-vx^0)$.

For static configurations the tension of the wall
\begin{equation}
\Omega(\Phi)=\lim_{l\to\infty}\frac{E[\Phi]}{l^{d-1}}=\int dx^1 \Big[ \frac{1}{2}\, \sum_{a=1}^N \frac{d \phi_a}{dx^1}\cdot \frac{d\phi_a}{dx^1}\, +\, U(\phi_1, \cdots , \phi_N)\Big] \, =\, \int dx^1 \, \, \omega (x^1) \quad  \label{wate}
\end{equation}
is a finite magnitude. Here $l^{d-1}$ is a normalizing volume in the $(x^2,\dots ,x^d)$ hyperplane. In particular we are interested in the cases $d=1$ and $d=3$. If $d=1$ these solutions are referred to as kinks and in this context the wall tension becomes the kink energy. If $d=3$ these solutions will be solitonic (thick) $2$-branes orthogonal to the $x^1$-axis. The previous finite tension requirement is fulfilled if and only if the asymptotic conditions
hold:
\begin{equation}
\lim_{x^1\to \pm \infty} \, \frac{d \Phi}{dx^1}(x^1)\, =\, 0\quad
,\qquad \lim_{x^1\to \pm \infty} \, \Phi(x^1)\, \in {\cal M} \,  .\label{asy}
\end{equation}
Therefore the domain walls connect asymptotically two vacua $\phi^{(i)}$ and $\phi^{(j)}$ of ${\cal M}$. The factorization of the potential energy density (\ref{susy01}) allows us to use the Bogomolny splitting of the wall tension. If the superpotential is a $C^2(\mathbb{R}^N)$-function along the integration path in $\mathbb{R}^N$ between two critical points of $W$, the so called BPS-domain walls are solutions of the first-order equations
\begin{equation}
\frac{d\phi_a}{dx^1} = \frac{\partial W}{\partial \phi_a} \hspace{0.5cm},\hspace{0.5cm} a=1,\dots,N \hspace{0.4cm},
\label{bps}
\end{equation}
and saturate the BPS bound, $\Omega_B[\Phi] = |W(\Phi^{c(i)})-W(\Phi^{c(j)})|$. If we know a static solution $\Phi(x^1,\dots,x^d)$ we can obtain a family of these solutions by means of the expression $\Phi(\overline{x}^1,x^2,\dots, x^d)$ where $\overline{x}^1 = (-1)^\beta \frac{x^1-a-{v}x^0}{\sqrt{1-{v}^2}}$ with $\beta=0,1$ and $a\in \mathbb{R}$ by simply using the symmetries of the model.

The normal modes of fluctuations around BPS domain wall solutions $\Phi_{\rm DW}(x^1)$ of (\ref{bps}), which are determined by the Laplace/Schr$\ddot{\rm o}$dinger type operator
\[
\mathbb{L}= -\nabla^2 \mathbf{I}_{N\times N} + \mathbf{v}^2 + \mathbf{V}(x^1) \hspace{0.3cm} \mbox{where} \hspace{0.3cm}  V_{ab}(x^1)=[\mathbf{V}(x^1)]_{ab}= \frac{\partial^2 U}{\partial \phi_a \partial \phi_b}[\Phi_{\rm DW}(x^1)] - v_a^2 \delta_{ab}
\]
have the form $\delta_\lambda \phi_a(x^0,\vec{x})=e^{i(\lambda+k_2^2+\dots + k_d^2) x^0}\xi_{a\lambda}(x^1)e^{-ik_2x^2-\dots-ik_nx^n}$. Here
the vectors $\Xi_\lambda(x^1)=\left( \xi_{1\lambda}(x^1) \cdots \xi_{N\lambda}(x^1) \right)^t$
are the eigenfunctions, $\mathbb{K} \,\Xi_\lambda(x^1) =\lambda\, \Xi_\lambda(x^1)$, of the second-order differential matrix operator:
\begin{equation}
\mathbb{K}= - \frac{\partial^2}{\partial(x^1)^2}\, \mathbf{I}_{N\times N} + \mathbf{v}^2 + \mathbf{V}(x^1) \hspace{0.4cm} . \label{operatorK}
\end{equation}
For domain walls interpolating between two vacua belonging to the same orbit of the (broken) symmetry group one finds an asymptotic behavior in the potential wells of the Schr$\ddot{\rm o}$dinger operator $\mathbb{K}$ of the form:
\[
\lim_{x \rightarrow \pm \infty} \mathbf{V} (x) = \mathbf{0}_{N\times N} \hspace{0.4cm},
\]
such that the behavior of the operator $\mathbb{K}$ asymptotically approaches to the free particle differential operator
\[
\mathbb{K}_0= - \frac{\partial^2}{\partial(x^1)^2}\, \mathbf{I}_{N\times N} + \mathbf{v}^2  \hspace{0.4cm} .
\]
Because the domain wall solutions break the spatial translational symmetry in the $x^1$-direction there is always a zero mode, a bound state of zero energy,
in the spectrum of $\mathbb{K}$. Other continuous symmetries broken by the domain wall mean that there are more zero modes up to a maximum number of $N$. In fact, it can be easily shown by deforming the first-order equations (\ref{bps}) that the operator (\ref{operatorK}) factorizes in the form $\mathbb{L}=\vec{\mathbb{A}}^\dagger \vec{\mathbb{A}}$ in terms of the first-order differential operator
\[
\vec{\mathbb{A}}=-\vec{\nabla} \,\, \textbf{I}_{N\times N} + D^2 W(\Phi_{DW}(x^1))
 \hspace{0.5cm} \mbox{where} \hspace{0.5cm} [D^2 W(\Phi_{DW}(x^1))]_{ab} = \frac{\partial^2 W}{\partial \phi_a \partial \phi_b}(\Phi_{\rm DW}(x^1))
\]
and its adjoint. Therefore, the zero modes $\Xi_{0\ell}$ are solutions of the system of first-order linear differential equations:
\begin{equation}
\vec{\mathbb{A}}\, \, \Xi_{0\ell}=0  \quad , \quad \ell=1,2, \cdots , \leq N \quad ,\label{edozero}
\end{equation}
which are thus the zero modes of $\mathbb{K}$ times the constant eigenfunctions of the operator $\mathbb{L}-\mathbb{K}$ (absence of transverse to the wall in the space plane waves).

\subsection{A model with a one-parametric family of iso-tension domain walls}

In what follows, we shall address the specific case where $N=2$ and the superpotential, depending also on a non-dimensional real parameter $\sigma$ that sets the strength of the coupling between the two scalar fields, is:
\[
W(\phi_1,\phi_2)= \frac{2}{3}\phi_1^3 -\frac{1}{2} \phi_1 + \sigma \phi_1\phi_2^2 \quad .
\]
The dynamics of this two-scalar field model is thus governed by the potential energy density:
\begin{equation}
U(\phi_1,\phi_2)= \frac{1}{2} \left(2 \phi_1^2 +\sigma \phi_2^2 -\frac{1}{2} \right)^2+2\sigma^2 \phi_1^2 \phi_2^2 \hspace{0.5cm}.
\label{bnrtpotential}
\end{equation}
Note that this function is a quartic polynomial in the fields and that the symmetry group of the system is discrete, the ${\rm G}=\mathbb{Z}_2 \times \mathbb{Z}_2$ generated by the field reflections: $\phi_1\to -\phi_1$ and $\phi_2\to -\phi_2$. There exist four critical points of $W(\Phi)$ if $\sigma\in\mathbb{R}^+$ is positive, namely:
\[
{\cal M}= \left\{\Phi^{c(1)}={\frac{1}{2}\choose 0}  \, , \, \Phi^{c(2)}=-{\frac{1}{2}\choose 0} \, , \, \Phi^{c(3)}={ 0 \choose \frac{1}{\sqrt{2\sigma}}} \, , \, \Phi^{c(4)}=-{ 0 \choose \frac{1}{\sqrt{2\sigma}}} \right\} \quad .
\]
The moduli space of vacua ${\cal M}/{\rm G}\simeq \{\Phi^{c(1)},\Phi^{c(3)}\}$ is thus formed by two points, whereas the symmetry is spontaneously broken to a $\mathbb{Z}_2$ subgroup (different in each point of the moduli) through the choice of vacuum to pass to the quantum theory. On each type of vacuum two meson branches emerge characterized respectively by the second-order vacuum fluctuation operators:
\[
\mathbb{L}_0^{(1)}=\mathbb{L}_0^{(2)}=\left( \begin{array}{cc} -\nabla^2+4 & 0 \\ 0 & -\nabla^2 +\sigma^2 \end{array} \right) \hspace{0.2cm},\hspace{0.2cm} \mathbb{L}_0^{(3)}=\mathbb{L}_0^{(4)}=\left( \begin{array}{cc} -\nabla^2+2\sigma & 0 \\ 0 & -\nabla^2 +2 \sigma \end{array} \right) \hspace{0.4cm}.
\]
In this model the Bogomolny equations (\ref{bps}) become:
\begin{equation}
\frac{d\phi_1}{dx^1} = \frac{\partial W}{\partial \phi_1} =  2\phi_1\phi_1+\sigma \phi_2\phi_2-\frac{1}{2}  \hspace{0.5cm},\hspace{0.5cm} \frac{d\phi_2}{dx^1} = \frac{\partial W}{\partial \phi_2} = 2\sigma \phi_1\phi_2 \hspace{0.5cm} .
\label{bps2}
\end{equation}
From (\ref{bps2}) we can obtain analytically a one-parametric family of orbits relating the field components $\phi_1$ and $\phi_2$ such that all of them correspond to BPS-domain walls:
\begin{eqnarray}
\Big(\frac{\gamma^2}{2\sigma} \Big)^\frac{1}{\sigma} \Big[4 \phi_1\phi_1 + \frac{2\sigma}{1-\sigma} \phi_2\phi_2 - 1 \Big] = \Big( \frac{\gamma^2}{1-\sigma}-1 \Big) (\phi_2\phi_2)^\frac{1}{\sigma} \hspace{1cm} &\mbox{if }& \sigma\neq 1 \hspace{0.4cm},  \nonumber \\
2\gamma^2 \phi_1\phi_1 + \phi_2\phi_2 \Big(1-\gamma^2 \log \frac{2\phi_2\phi_2}{\gamma^2} \Big) = \frac{1}{2}\gamma^2 \hspace{1cm} &\mbox{if }& \sigma= 1 \hspace{0.4cm}. \label{orbitkink}
\end{eqnarray}
The integration constant $\gamma$ has been arranged in such a way that the finite-tension domain walls are given by the orbits in the range $\gamma\in [0,1)$, a range which is independent of the coupling constant $\sigma$. All the $\Phi_{\rm DW}(x^1;\gamma)$-orbits, $\gamma\in [0,1)$, connect the vacuum points $\Phi^{c(1)}$ and $\Phi^{c(2)}$, see Figure 1. Remarkably, the wall tension of all these BPS-domain walls is the same:
\[
\Omega \left[\Phi_{\rm DW}(x^1);\gamma\right]= \left|W(\Phi^{c(1)}) - W(\Phi^{c(2)})\right| = \frac{1}{3} \hspace{0.4cm} , \, \,  \forall\gamma\in[1,0) \, \, .
\]
The geometric meaning of $\gamma$ is clear: it determines the point of the BPS-orbits where the curves cross the $\phi_2$-axis (see Figure 1). From the first equation in (\ref{orbitkink}) we check that the solution for $\phi_1(x_0^1,\gamma)=0$
is: $\bar\phi_2(x_0^1,\gamma)=\frac{\gamma}{\sqrt{2\sigma}}$.

\begin{figure}[h]
\centerline{\includegraphics[height=5cm]{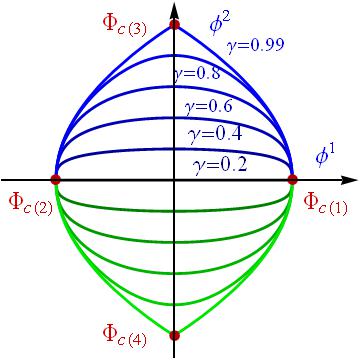}}
\caption{\small $\Phi_{\rm DW}(x^1,\gamma)$-orbits for $\sigma=\frac{1}{2}$ and several values of the parameter $\gamma\in [0,1)$
including $\gamma=1$ in the boundary curves.}
\end{figure}

The choice of the $\gamma=0$ point in the moduli space of BPS-walls corresponds to a particularly simple solution
\begin{equation}
\phi_1\vert_{\rm DW}(x^1;0)=\frac{1}{2} \tanh \overline{x}^1 \quad , \quad \phi_2\vert_{\rm DW}(x^1;0)=0 \quad , \label{tw1}
\end{equation}
living on the $[-1,1]$-interval of the $\phi_1$-axis, see Figure 1.
The energy per unit of volume of this wall, the integrand in (\ref{wate}), is concentrated around the point $\overline{x}^1=0$, see Figure 3: $\omega(x^1)=\frac{1}{4}{\rm sech}^4 \overline{x}^1$. The value $\gamma=1$ of the integration constant gives rise to domain walls which do not belong to the the same topological sector as all the others in the BPS $\gamma$-family but live on the critical orbits joining different points
inthe vacuum moduli space, see Figure 1. Moreover, their wall tension is half the tension of the BPS-walls:
\[
\Omega \left[\Phi_{\rm DW}(x^1);1\right]= \left|W(\Phi^{c(1)}) - W(\Phi^{c(3)})\right| = \frac{1}{6} \hspace{0.4cm}  \, \, .
\]

\begin{figure}[ht]
\centerline{
\includegraphics[width=3.2cm]{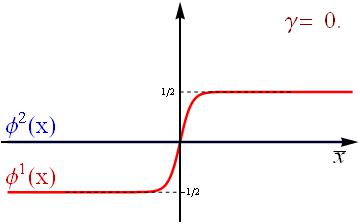}\hspace{0.5cm}
\includegraphics[width=3.2cm]{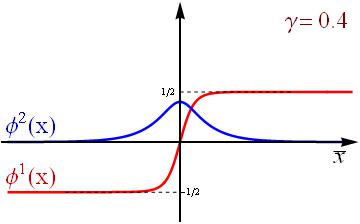}\hspace{0.5cm}
\includegraphics[width=3.2cm]{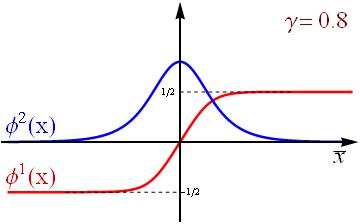}\hspace{0.5cm}
\includegraphics[width=3.2cm]{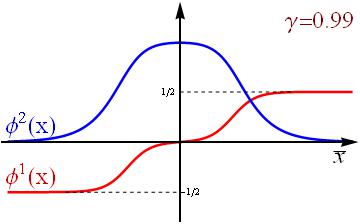}}

\caption{\small Domain wall profiles $\Phi_{\rm DW}^{\sigma=\frac{1}{2}}(\overline{x}^1,\gamma)$ for several values of $\gamma$.}
\end{figure}

In general, it is not possible to obtain analytical expressions for the BPS-domain wall profiles. The generic profiles depending in $\gamma$ for any $\sigma$ follow the pattern displayed in Figure 2; they interpolate between $\Phi_{\rm DW}(x^1;0)$ and two $\Phi_{\rm DW}(x^1;1)$-configurations very far apart. Thus, these BPS-domain walls reveal a composite structure as $\gamma$ approaches to the critical value 1. Changes in the domain wall profiles when $\gamma$ approaches $1$ are localized around two points. This fact suggests that the $\Phi_{\rm DW}(\overline{x}^1,\gamma)$ solutions are composed of two constituent domain walls, an evident proposition shown in the plots of the $\Phi_{\rm DW}(\overline{x}^1,\gamma)$ energies per unit of volume, see Figure 3. For $\gamma=0.99$ two identical lumps of energy per volume unit located at two distant points on the $x^1$-axis arise. By contrast, at small values of $\gamma$ the two lumps appear on top of each other. In sum, the $\Phi_{\rm DW}(\overline{x}^1,\gamma)$ walls can be thought of as a non-linear combination of two basic identical extended objects separated by a certain distance (non-linearly) measured by $\gamma$. At the classical level, the basic objects experience no repulsive or attractive forces between each other; they move freely in the moduli space of solutions of the first-order ODE system (\ref{bps2}) parametrized by $a$ and $\gamma$. $a$ describes the center of mass of the two basic walls, whereas $\gamma$ is the relative coordinate between them. There is no preferred separation $\gamma$ between the constituent lumps.

\begin{figure}[ht]
\centerline{\includegraphics[width=3.2cm]{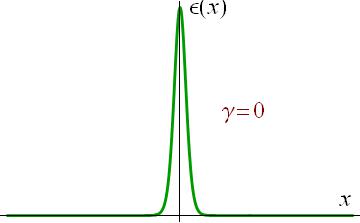}\hspace{0.5cm}
\includegraphics[width=3.2cm]{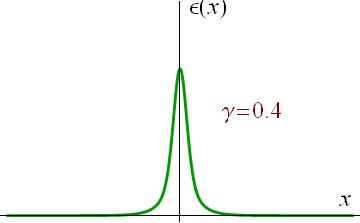}\hspace{0.5cm}
\includegraphics[width=3.2cm]{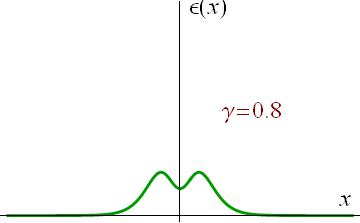}\hspace{0.5cm}
\includegraphics[width=3.2cm]{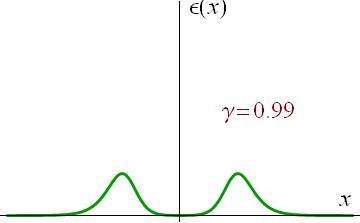}}
\caption{\small Generic behavior of the $\Phi_{\rm DW}(\overline{x}^1;\gamma)$ domain wall energy per volume unit for several values of the family parameter  $\gamma$.}
\end{figure}

The adiabatic scattering of these composite domain walls has been studied in \cite{Alonso2002b} within Manton's principle of geodesic motion  in the BPS moduli space equipped with the metric inherited from the zero modes: $\frac{\partial\Phi_{\rm DW}}{\partial a}$ and $\frac{\partial\Phi_{\rm DW}}{\partial \gamma}$, see \cite{Manton2004}. The classical energy per unit of surface (tension) degeneracy of the $\Phi_{\rm DW}(\overline{x}^1,\gamma)$ BPS-walls prompts a natural question: do the quantum fluctuations rule out the classical degeneracy of the two twin lumps located at any distance with respect to each other?  In other words, does a quantum phase transition take place in this system, inducing forces between the constituent lumps of wall tension? This issue will be the main concern of the rest of the paper,  after developing in Section 3 a modification of the standard Gilkey-de Witt heat kernel expansion designed to cope with the problems posed by infrared divergences (zero modes). The improved procedure will produce an estimation of the one-loop shift of domain wall tension that is precise enough to answer this question in a remarkable outcome.

\section{Heat kernel asymptotic expansion for an ordinary differential operator with zero modes}

In this Section we shall generalize the improved heat kernel expansion developed in \cite{Alonso2012} designed to cope with zero mode fluctuations to models in $N$-scalar field theories. We shall present this new expansion in $(1+1)$-dimensions because the topological wall defects to be addressed depends only on the $x_1$-coordinate. It is not only that the results in this Section will be appliable to calculate both kink mass
and wall tension shifts. The suitable modification in the heat kernel expansion when zero modes exist in the spectrum of defect fluctuations
is easy to grasp in one spatial dimension.

The standard Gilkey-de Witt heat kernel expansion works fine for operators with a strictly positive spectrum. In this class of systems the Gilkey-DeWitt procedure is very effective for attacking problems where the ultraviolet part of the spectrum plays a prominent r$\hat{\rm o}$le: calculations of anomalies at one-loop order, resummation of fluctuations on constant backgrounds described by effective actions, etcetera, see \cite{Vassilevich2003}. Fluctuations around extended objects, however, always give rise to zero modes, e.g., disguised as kinks or domain walls the BPS solutions of our model show two null fluctuation modes: $\Xi_{01}(\overline{x})=\frac{\partial\Phi_{\rm DW}(\overline{x};\gamma)}{\partial\overline{x}}$, $\Xi_{02}(\overline{x})=\frac{\partial\Phi_{\rm DW}(\overline{x};\gamma)}{\partial\gamma}$. Thus, the infrared effects become important, especially because these effects are not tamed by the subtraction of the fluctuations around the vacuum that do not show infrared problems. The Gilkey-DeWitt heat trace expansion must to be modified to accommodate the impact of zero modes  and we shall obtain the improved expansion by generalizing the ideas described in references \cite{Alonso2011,Alonso2012,Alonso2013b} to $N$-component scalar field-theory models, one of the main theoretical novelties of this paper.

Let $\mathbb{K}$ be a general ordinary differential matrix operator of the general form shown in (\ref{operatorK}). The spectral $\mathbb{K}$-heat trace $h_\mathbb{K}(\beta)={\rm Tr}_{L^2(\mathbb{S}^1)}\, e^{-\beta \mathbb{K}}$ admits an integral kernel representation
\begin{equation}
h_\mathbb{K}(\beta)=\int_\mathbb{R} \, dx \, {\rm tr}\, \mathbf{K}_\mathbb{K}(x,x;\beta) \, \,
\label{heattrace}
\end{equation}
where ${\rm tr}$ stands for trace in the matricial sense. The spectral decomposition
of the matrix heat kernel in terms of the bound state and scattering eigenfunctions reads:
\begin{equation}
\mathbf{K}_{\mathbb{K}}(x,y;\beta)= \sum_{\ell=1}^{N_{zm}} \Xi_{0\ell}(x)\, \Xi_{0\ell}^\dagger (y)+\sum_{n=1}^{N_B} \Xi_n(x) \, \Xi_n^\dagger(y) e^{-\beta \omega_n^2}+ \int \! dk \, \Xi_k (x) \, \Xi_k^\dagger(y) \, e^{-\beta \omega^2(k)} \hspace{0.4cm}.
\label{integralkernel}
\end{equation}
Here $N_{zm}$ denotes the number of zero modes $\Xi_{0\ell}(x)$, linearly independent functions in the algebraic kernel of $\mathbb{K}$, $N_B$ is the number of bound states, $\Xi_n(x)$, in the positive spectrum of $\mathbb{K}$, and $\Xi_k(x)$ are the continuous spectrum eigenfunctions of the kink fluctuation matrix operator $\mathbb{K}$. $\Xi_{0\ell}(x)$, $\Xi_n(x)$ and $\Xi_k(x)$ are $N$-component column vectors and form an orthonormal basis in the Hilbert space. The key observation is that the zero mode contribution is $\beta$-independent because the eigenvalue of a zero mode vanishes.

The matrix heat kernel (\ref{integralkernel}) is the fundamental solution of the $\mathbb{K}$-heat equation:
\begin{equation}
\left(\frac{\partial}{\partial\beta}+\mathbb{K}\right) \mathbf{K}_{\mathbb{K}}(x,y;\beta)=0 \hspace{0.5cm} , \hspace{0.5cm} \mathbf{K}_\mathbb{K}(x,y;0)=\delta(x-y)  \mathbf{I}_{N\times N} \hspace{0.4cm}, \label{heateq}
\end{equation}
becoming a Dirac delta distribution at infinite temperature $\beta=0$. The asymptotic behavior of $\mathbf{K}_\mathbb{K}(x,y;\beta)$ at zero temperature $\beta=+\infty$ is, however, determined from the zero modes:
\begin{equation}
\lim_{\beta\rightarrow +\infty} \mathbf{K}_\mathbb{K}(x,y;\beta)= \sum_{\ell=1}^{N_{zm}}\Psi_{0\ell}(x) \Psi_{0\ell}^\dagger(y) \hspace{0.4cm} .
\label{asympintkernel}
\end{equation}
The Gilkey-DeWitt procedure profits from knowledge of the $\mathbb{K}_0$-heat kernel
\begin{equation}
\mathbf{K}_{\mathbb{K}_0}(x,y;\beta)=\frac{e^{- \frac{(x-y)^2}{4 \, \beta\,}} }{\sqrt{4\pi \beta}} \, e^{-\beta \mathbf{v}^2} \hspace{0.5cm},\hspace{0.5cm} e^{-\beta \mathbf{v}^2} = {\rm diag}(e^{-\beta v_1^2},\dots, e^{-\beta v_N^2}) \hspace{0.4cm},
\label{k0}
\end{equation}
by assuming a factorization of the $\mathbb{K}$-heat kernel in the form
\begin{equation}
\mathbf{K}_\mathbb{K}(x,y;\beta) = \mathbf{A}(x,y;\beta) \mathbf{K}_{\mathbb{K}_0}(x,y;\beta) \hspace{0.4cm}, \label{standardfact}
\end{equation}
and solving the subsequent transfer equation for $\mathbf{A}(x,y;\beta)$ as a power series in $\beta$ with the infinite temperature limit $\mathbf{A}(x,y;0)=\mathbf{I}_{N\times N}$ because $\mathbf{K}_{\mathbb{K}_0}(x,y;0)=\delta(x-y)\mathbf{I}_{N\times N}$.
The low temperature limit deduced from (\ref{k0})
\[
\lim_{\beta\rightarrow +\infty} \mathbf{K}_{\mathbb{K}_0}(x,y,\beta) =0\hspace{0.4cm},
\]
produces a mismatch with the low temperature value of $\mathbf{K}_\mathbb{K}(x,y;\beta)$ determined from (\ref{integralkernel})
if zero modes are present, such that the standard factorization (\ref{standardfact}) fails at low temperature. Therefore, one expects departures from the exact value of $h_\mathbb{K}(\beta)$ for $\beta$ large enough in the computation from the Gilkey-DeWitt-Avramidi high-temperature expansion. To escape this problem we propose a new factorization
\begin{equation}
\mathbf{K}_\mathbb{K}(x,y;\beta) = \mathbf{C}(x,y;\beta) \mathbf{K}_{\mathbb{K}_0}(x,y;\beta) + \sum_{\ell=1}^{N_{zm}} e^{-\frac{(x-y)^2}{4\beta}} \Xi_{0\ell}(x)\Xi_{0\ell}^\dagger(y) \mathbf{G}_\ell(\beta)  \label{newfact}
\end{equation}
as the basic assumption to implement the Gilkey-DeWitt heat kernel expansion.
The matrix kernel $\mathbf{C}(x,y;\beta)$ behaves as demanded by the kernel at infinite temperature, whereas the as yet unspecified  matrix function $\mathbf{G}(\beta)$ which accompanies the zero modes must be chosen with the unique criterion of reproducing (\ref{newfact}) the right behavior of the $\mathbb{K}$-heat kernel at both high and low temperatures in the new factorization. Requiring
\begin{equation}
\lim_{\beta\to 0}\,\mathbf{C}(x,y;\beta)=\mathbf{I}_{N\times N} \hspace{0.5cm},\hspace{0.5cm} \lim_{\beta\to 0} \, \mathbf{G}_\ell(\beta)=\mathbf{0}_{N\times N} \hspace{0.5cm},\hspace{0.5cm} \lim_{\beta\rightarrow +\infty} \mathbf{G}_\ell(\beta) =\mathbf{I}_{N\times N} \label{cond14}
\end{equation}
the asymptotic behavior deduced from (\ref{integralkernel}) is ensured at both limits. It is obvious that suppression of the zero modes $\Psi_{0\ell}(x)$ in (\ref{newfact}) reproduces the standard factorization (\ref{standardfact}), such that the matrix kernel $\mathbf{C}(x,y;\beta)$ becomes the matrix kernel $\mathbf{A}(x,y;\beta)$. From now on we follow a fairly standard path supplemented by an appropriate choice of $\mathbf{G}_\ell(\beta)$. First, $\mathbf{C}(x,y;\beta)$ is expanded as a power series on the variable $\beta$
\begin{equation}
\mathbf{C}(x,y;\beta)=\sum_{n=0}^\infty \mathbf{c}_n(x,y)\beta^n \hspace{0.5cm},\hspace{0.5cm} \mathbf{c}_0(x,y)=\mathbf{I}_{N\times N} \hspace{0.4cm}.
\label{series01}
\end{equation}
Second, the matrix function $\mathbf{G}_\ell(\beta)$ is chosen from the error function:
\begin{equation}
\mathbf{G}_\ell(\beta)= {\rm erf}(\mathbf{v} \sqrt{\beta}) \hspace{0.5cm},\hspace{0.5cm}  {\rm erf}(\mathbf{v} \sqrt{\beta})= {\rm diag}[{\rm erf}(v_1\sqrt{\beta}),\dots,{\rm erf}(v_N\sqrt{\beta})] \hspace{0.4cm}.
\label{g01}
\end{equation}
There is a first, and obvious, reason for this choice: (\ref{g01}) implies (\ref{cond14}). A second, hidden, reason arises
when we plug
\begin{equation}
\mathbf{K}_\mathbb{K}(x,y;\beta) = \sum_{n=0}^\infty \frac{e^{- \frac{(x-y)^2}{4 \, \beta\,}} }{\sqrt{4\pi}} \beta^{n-\frac{1}{2}} \mathbf{c}_n(x,y) \, e^{-\beta \mathbf{v}^2} + \sum_{\ell=1}^{N_{zm}} e^{-\frac{(x-y)^2}{4\beta}} \Xi_{0\ell}(x)\Xi_{0\ell}^\dagger(y) {\rm erf}(\mathbf{v} \sqrt{\beta})  \label{newfact2}
\end{equation}
into the heat equation (\ref{heateq}).  The recurrence relations
\begin{eqnarray}
&&(n+1)\mathbf{c}_{n+1}(x,y) +(x-y)\frac{\partial \mathbf{c}_{n+1}(x,y)}{\partial x} - \frac{\partial^2 \mathbf{c}_n(x,y)}{\partial x^2} + \mathbf{V}(x)\mathbf{c}_n(x,y) + \nonumber \\
&& + [\mathbf{v}^2,\mathbf{c}_n(x,y)]+\sum_{\ell=1}^{N_{zm}} \Big[ 2 \, \Xi_{0\ell}(x) \Xi_{0\ell}^\dagger(y) \mathbf{v} \delta_{0n} + \frac{2^{n+1}}{(2n+1)!!}  \Xi_{0\ell}(x) \Xi_{0\ell}^\dagger(y) \mathbf{v}^{2n+1} + \label{recurrence1} \\ && + \frac{2^{n+2}(x-y)}{(2n+1)!!} \frac{d \, \Xi_{0\ell}(x)}{dx} \Xi_{0\ell}^\dagger(y)  \mathbf{v}^{2n+1}\Big]=0 \nonumber
\end{eqnarray}
between the densities $\mathbf{c}_n(x,y)$ and their derivatives must be solved. Besides providing the right behavior at high and low temperatures (\ref{g01}) the choice $\mathbf{G}_\ell(\beta)= {\rm erf}(\mathbf{v} \sqrt{\beta})$ minimizes the difficulties  in solving the recurrences (\ref{recurrence1}). One might interpret the factorization (\ref{newfact}) as being based on a background breaking the same symmetries as the extended object (giving rise to the zero modes) and the choice of the error function would correspond to the simplest background prompting the same symmetry breaking.

The calculation of the  $\mathbb{K}$-heat trace (\ref{heattrace}) needs to use only the diagonal densities. The very delicate limit $y\to x$ must be taken in (\ref{newfact2})
\begin{equation}
\mathbf{K}_\mathbb{K}(x,x;\beta) = \sum_{n=0}^\infty \frac{\beta^{n-\frac{1}{2}}}{\sqrt{4\pi}} \,\, {}^{(0)} \mathbf{C}_n(x) \, e^{-\beta \mathbf{v}^2} + \sum_{\ell=1}^{N_{zm}} |\,\Xi_{0\ell}(x)|^2 {\rm erf}(\mathbf{v} \sqrt{\beta}) \, . \label{newfact3}
\end{equation}
The identification of the densities $\mathbf{C}_n(x)$ also requires the implementation of the limit $y\rightarrow x$ in the recurrence relations (\ref{recurrence1}). We shall use the notation
\begin{equation}
{}^{(0)} \mathbf{C}_n(x)= \lim_{y\rightarrow x} \mathbf{c}_n(x,y) \hspace{0.8cm},\hspace{0.8cm}
{}^{(k)} \mathbf{C}_n(x) =\lim_{y\rightarrow x} \frac{\partial^k \mathbf{c}_n(x,y)}{\partial x^k}\hspace{0.4cm},
\label{coefaux}
\end{equation}
as a practical tool to solve the recurrence relations
\begin{eqnarray}
&& (n+k) {}^{(k)}\mathbf{C}_n(x)= {}^{(k+2)}\mathbf{C}_{n-1}(x) -\sum_{j=0}^k {k\choose j} \frac{\partial^j \mathbf{V}(x)}{\partial x^j} {}^{(k-j)}\mathbf{C}_{n-1}(x)- [\mathbf{v}^2,{}^{(k)}\mathbf{C}_{n-1}(x)] - \nonumber \\
&& -\sum_{\ell=1}^{N_{zm}} \Big[ 2\frac{\partial^k \,\Xi_{0\ell}(x)}{\partial x^k} \,\Xi_{0\ell}^\dagger(x) \mathbf{v} \delta_{0,n-1} + (1+2k) \frac{2^n}{(2n-1)!!} \frac{\partial^k \,\Xi_{0\ell}(x)}{\partial x^k}\, \Xi_{0\ell}^\dagger(x) \mathbf{v}^{2n-1} \Big] \hspace{0.4cm} ,\label{recurrence2}
\end{eqnarray}
where $\mathbf{v}={\rm diag}(v_1, v_2, \cdots , v_N)$, starting from
\[
{}^{(k)} \mathbf{C}_0(x)=\delta_{0k}\, \mathbf{I}_{N\times N} \, \, .
\]
These latter recurrence relations have been derived by taking the $k$-th derivative in (\ref{recurrence1}) with respect to the spatial variable $x$ and then taking the limit when the $y$ variable approaches $x$. This ordering in the (mutually non-commuting) operations of taking derivatives with respect to $x$ first and going to the $y\to x$ diagonal limit later in (\ref{recurrence1}) is explicitly implemented in the notation shown in (\ref{coefaux}).
We show the first three densities obtained from the recurrences (\ref{recurrence2})
\begin{eqnarray*}
{}^{(0)}\mathbf{C}_0(x)&=&\mathbf{I}_{N\times N}\hspace{0.4cm}, \\
{}^{(0)}\mathbf{C}_1(x)&=& -\mathbf{V}(x)- \sum_{\ell=1}^{N_{zm}} 4\, \Xi_{0\ell}(x)\, \Xi_{0\ell}^\dagger(x) \mathbf{v} \hspace{0.4cm},\\
{}^{(0)}\mathbf{C}_2(x)&=& - \frac{1}{6} \mathbf{V}''(x) + \frac{1}{2} \mathbf{V}^2(x) + \frac{1}{2} [\mathbf{v}^2,\mathbf{V}(x)] - \frac{8}{3} \sum_{\ell=1}^{N_{zm}} \Xi_{0\ell}(x) \,\Xi_{0\ell}^\dagger \mathbf{v}^3\hspace{0.4cm},
\end{eqnarray*}
needed in (\ref{newfact3}) to determine the matrix heat kernel on the diagonal $x=y$. We must compute the Seeley coefficients
\begin{equation}
c_n^a(\mathbb{K})=\int dx \, \, [{}^{(0)}\mathbf{C}_n(x)]_{aa} \label{modifiedSeeley}
\end{equation}
and some other new ones coming from the zeros modes
\begin{equation}
f^a_\ell(\mathbb{K}) = \int dx \, \, \left| [\,\Xi_{0\ell}(x)]_a \right|^2 \label{modifiedf}
\end{equation}
which arises by taking the matrix trace and integrating over the real line the different summands in (\ref{newfact3}) to find the series expansion of the $\mathbb{K}$-heat trace $h_\mathbb{K}(\beta)$. Subtraction of the $h_{\mathbb{K}_0}(\beta)$ heat function suppresses the contribution of the $c_0^a(\mathbb{K})$ coefficients. Finally, the asymptotic series formula reads
\begin{equation}
h_\mathbb{K}(\beta)-h_{\mathbb{K}_0}(\beta)=\sum_{n=1}^\infty \sum_{a=1}^N c_n^a(\mathbb{K})\,  e^{-\beta v_a^2} \frac{1}{\sqrt{4\pi}} \beta^{n-\frac{1}{2}} + \sum_{\ell=1}^{N_{zm}} \sum_{a=1}^N f^a_\ell(\mathbb{K}) \, {\rm erf}(v_a \sqrt{\beta}) \hspace{0.2cm} .
\label{heattrace2}
\end{equation}
The Seeley coefficients of first-order in (\ref{heattrace2})
\begin{equation}
c_1^a(\mathbb{K}) = -\int\! dx \, V_{aa}(x) - 4 v_a \sum_{\ell=1}^{N_{zm}} f^a_\ell(K)
\label{firstseeley}
\end{equation}
differ from the standard ones only in the zero mode contribution.

\section{One-loop mass shifts of the $\Phi_{\rm TK}(\overline{x};\gamma)$ degenerate kinks}

In this Section we shall apply the general formulas obtained in Section \S.3 to estimate the shifts induced by one-loop fluctuations in the mass of any member of the degenerate family of BPS solutions described in the Section \S.2 2 for the particular case $d=1$. Thus, we denote the BPS kink defects as $\Phi_{\rm TK}(\overline{x};\gamma)$ where we recall that $\gamma\in [0,1)$ is the parameter caracterizing the distance between the two constituent kins. The main issue is to investigate whether or not the shifts depend on $\gamma$. Dependence of the kink mass shifts on $\gamma$ would imply that attractive or repulsive forces arise between the constituent kinks due to one-loop kink fluctuations in such a way
that the defects in the family cease to be BPS.

On attempting to accomplish this task, two difficulties arise that we must comment on before of solving these problems in turn. We recall that the efficiency of the DHN procedure depends critically on a complete knowledge of the spectral data of the operator $\mathbb{K}$: bound state eigenvalues and scattering wave phase shifts \cite{Alonso2002c}. Except for $\gamma=0$, the second-order kink fluctuation operator is a non-diagonal matrix differential operator. The identification of the bound state eigenvalues and the phase shifts is impossible in all these $\gamma\neq 0$ cases. Recall that $\mathbb{K}[\Phi_{\rm TK}(\overline{x};0)]$ is not only diagonal but that the spectral problems of the diagonal Schr$\ddot{\rm o}$dinger operators are exactly solvable. In order to circumvent the lack of spectral information when $\gamma\neq 0$ we shall develop the following strategy: (1) We shall use the spectral zeta function regularization method to control the ultraviolet divergences arising in the computation of one-loop kink mass shifts. (2) The spectral zeta function will be determined from the Mellin transform of the heat trace of the operator $\mathbb{K}[\Phi_{\rm TK}(\overline{x};\gamma)]$. (3) The $\mathbb{K}$-heat trace will be evaluated, even without knowing the details of the spectrum of $\mathbb{K}$, by means of the Gilkey-de Witt heat kernel asymptotic expansion.

\subsection{Spectral zeta function regularization and one-loop kink mass shifts}

Use of the heat kernel/zeta function based on the modified GDW expansion in the computation of one-loop kink mass shifts in $N$-component
scalar field models is briefly explained in this subsection. Formally, the kink Casimir energy is the difference between the $L^2$-traces of the fluctuation operators around the kink and the vacuum:
\begin{equation}
\Delta \widetilde{E}_1[\Phi_{\rm TK}(x;\gamma)]=\frac{\hbar m}{2}\left\{{\rm Tr}_{\rm L^2}\, \mathbb{K}^{\frac{1}{2}}[\Phi_{\rm TK}(x;\gamma)]-{\rm Tr}_{\rm L^2}\, \mathbb{K}_0^{\frac{1}{2}}[\Phi^{\rm c(1)}] \right\}\label{dwct} \, .
\end{equation}
We start by regularizing the second summand in (\ref{dwct}), the vacuum energy induced by quantum fluctuations, by means of the spectral zeta function of $\mathbb{K}_0$. The value of the spectral zeta function of the $\mathbb{K}_0$-operator (a meromorphic function) at a regular point in $s\in{\mathbb C}$ is assigned to it:
{\small\begin{equation}
\frac{\hbar m}{2}{\rm Tr}_{\rm L^2}\, \mathbb{K}_0^{\frac{1}{2}}[\Phi^{\rm c(1)}]=\frac{\hbar m}{2}\zeta_{\mathbb{K}_0}(-{\textstyle\frac{1}{2}}) \, \, \rightarrow \, \, \frac{\hbar m}{2}\left(\frac{\mu^2}{m^2}\right)^{s+\frac{1}{2}}{\rm Tr}_{\rm L^2}\, \mathbb{K}_0^{-s}[\Phi^{\rm c(1)}]=\frac{\hbar m}{2}\left(\frac{\mu^2}{m^2}\right)^{s+\frac{1}{2}}\zeta_{K_0}(s)
\end{equation}}
where $\mu$ is a parameter of dimensions $L^{-1}$ introduced to keep the dimensions of the regularized energy right. We stress that a pole of this meromorphic function sits at the physical value $s=-\frac{1}{2}\in\mathbb{C}$. The same regularization procedure is applied to control the ultraviolet divergences due to kink fluctuations, i.e., the spectral zeta function of $\mathbb{K}$ is used to regularize the other summand in (\ref{dwct}). Thus, the kink Casimir energy is regularized in the form:
\begin{equation}
\Delta  \widetilde{E}_1[\Phi_{\rm TK}(x;\gamma)][s]= \frac{\hbar m}{2}\left(\frac{\mu^2}{m^2}\right)^{s+\frac{1}{2}}\left(\zeta_{\mathbb{K}}(s)-\zeta_{\mathbb{K}_0}(s)\right) \hspace{0.4cm}. \label{zrkc}
\end{equation}
The $\mathbb{K}$-zeta function is related to the $\mathbb{K}$-heat trace by means of a Mellin transform
\begin{equation}
\zeta_{\mathbb{K}}(s)=\frac{1}{\Gamma(s)} \int_0^\infty \, d\beta \, \beta^{s-1} \, h_{\mathbb{K}}(\beta) \hspace{0.3cm}, \label{Mellin54}
\end{equation}
such that the regularized shift in the kink mass (\ref{zrkc}) can be given in terms of the $\mathbb{K}$- and $\mathbb{K}_0$-heat traces:
\begin{equation}
\Delta  \widetilde{E}_1(\Phi_{\rm DW})[s]=\frac{\hbar m}{2} \left(\frac{\mu^2}{m^2}\right)^{s+\frac{1}{2}}\frac{1}{\Gamma(s)}\left[\int_0^\infty \, d\beta \, \beta^{s-1}\left(h_{\mathbb{K}}(\beta)-h_{\mathbb{K}_0}(\beta)\right)\right]\hspace{0.4cm}.
\label{zrkc2}
\end{equation}
Plugging the modified heat trace expansion (\ref{heattrace2}) into the Mellin transform (\ref{Mellin54}) we obtain
\[
\zeta_\mathbb{K}(s)-\zeta_{\mathbb{K}_0}(s) = \frac{1}{\sqrt{4\pi}} \sum_{n=1}^\infty \sum_{a=1}^N c_n^a(\mathbb{K})\, v_a^{1-2n-2s}\, \frac{\Gamma[s+n-\frac{1}{2}]}{\Gamma[s]} - \frac{1}{\sqrt{\pi}} \sum_{\ell=1}^{N_{zm}} \sum_{a=1}^N f_\ell^a(\mathbb{K}) \, v_a^{-2s} \, \frac{\Gamma[s+\frac{1}{2}]}{s\Gamma[s]} \,
\]
which provides the regularized kink Casimir energy in the form of the series:
\begin{eqnarray}
\Delta \widetilde{E}_1[\Phi_{\rm TK}](s) &=& \frac{\hbar m}{\sqrt{4\pi}} \Big( \frac{\mu^2}{m^2}\Big)^{s+\frac{1}{2}} \Big[ \frac{1}{2} \sum_{n=1}^\infty \sum_{a=1}^N c_n^a(\mathbb{K})\, v_a^{1-2n-2s}\, \frac{\Gamma[s+n-\frac{1}{2}]}{\Gamma[s]} - \nonumber \\ && - \sum_{\ell=1}^{N_{zm}} \sum_{a=1}^N f_\ell^a(\mathbb{K}) \, v_a^{-2s}\, \frac{\Gamma[s+\frac{1}{2}]}{s\Gamma[s]} \Big] \, .
\label{rest1}
\end{eqnarray}
The energy due to the one-loop mass renormalization counter-term can be also regularized in terms of the $\mathbb{K}_0$-zeta function as:
\begin{eqnarray}
\Delta  \widetilde{E}_2(\Phi_{\rm TK})[s]&=&  \frac{\hbar m}{2}\left(\frac{\mu^2}{m^2}\right)^{s+\frac{1}{2}}\lim_{L\to\infty}\frac{1}{L}\frac{\Gamma(s+1)}{\Gamma(s)}\sum_{a=1}^N\, \langle V_{aa}\rangle \,\zeta_{\mathbb{K}_{0aa}}(s+1) \nonumber \\ &=& \frac{\hbar m}{2}\left(\frac{\mu^2}{m^2}\right)^{s+\frac{1}{2}}\frac{\Gamma(s+\frac{1}{2})}{\Gamma(s)}\sum_{a=1}^N\, \frac{\langle V_{aa}\rangle}{v_a^{2s+1}}\hspace{0.2cm}.
\label{zrkc3}
\end{eqnarray}
Finally, we write the zeta function-regularized one-loop mass shift formula:
\begin{equation}
\Delta  \widetilde{E}(\Phi_{\rm TK})= \lim_{s\rightarrow -\frac{1}{2}} \Delta \widetilde{E}_1(\Phi_{\rm TK})[s]+ \lim_{s\rightarrow -\frac{1}{2}} \Delta \widetilde{E}_2(\Phi_{\rm TK})[s] \hspace{0.4cm}.
\label{zrkc4}
\end{equation}
A crucial cancelation, obeying the heat kernel renormalization criterion, occurs after the addition of these two contributions to the regularized one-loop kink mass shift $\Delta \widetilde{E}[\Phi_{\rm TK}](s) =\Delta \widetilde{E}_1[\Phi_{\rm TK}](s)+ \Delta \widetilde{E}_2[\Phi_{\rm TK}](s)$. $\Delta \widetilde{E}_2[\Phi_{\rm TK}](s)$ is annihilated by the part of the $n=1$ summands in (\ref{rest1}) that depend on $\langle V_{aa} \rangle$ entering in the first-order coefficients $c_1^a(\mathbb{K})$, see (\ref{firstseeley}). This cancelation is identical to the cancelation that occurs in the standard method and is very well known in the literature, see \cite{Bordag2002}. The novelty here is that by using the modified GDW expansion two divergences still remain at $s=-\frac{1}{2}${\footnote{The heat kernel renormalization criterion, which in $(1+1)$-dimensions is tantamount to the minimal renormalization achieved by normal ordering or the vanishing tadpole rule, applies only to massive fluctuations, not to the zero modes.}}. We now show that the residua at the poles at the physical point $s=-\frac{1}{2}$ due to the zero mode additions in the first Seeley coefficients and the last term of the Mellin transform (\ref{rest1}) are such that these divergences do not exactly cancel
\begin{eqnarray*}
&& \lim_{s\rightarrow -\frac{1}{2}} \left[ - \frac{2}{\sqrt{\pi}\,v_a^{2s} } \frac{\Gamma[s+\frac{1}{2}]}{\Gamma[s]} - \frac{1}{\sqrt{\pi}v_a^{2s}} \frac{\Gamma[s+\frac{1}{2}]}{s\Gamma[s]} \right] = \lim_{\varepsilon\rightarrow 0} \Big[ \frac{v_a}{\pi} \left(\frac{1}{\varepsilon} -\,[\gamma +2 \log v_a+ \psi(-{\textstyle\frac{1}{2}})] \right)+ \\
&& + {\cal O}_1(\varepsilon) -\frac{v_a}{\pi} \left(\frac{1}{\varepsilon} -[\gamma +2 \log v_a+ \psi(-{\textstyle\frac{1}{2}})]+2 \right)+ {\cal O}_2(\varepsilon) \Big] = -\frac{2v_a}{\pi} \quad ,
\end{eqnarray*}
but leave the finite remainder: $-\sum_{a=1}^N \frac{\hbar m v_a}{\pi} \sum_{\ell=1}^{N_{zm}} f_\ell^a(\mathbb{K})$. The one-loop correction to the classical kink mass obtained in the framework of the modified Gilkey-DeWitt heat kernel asymptotic is formulated as the truncated series:
\begin{equation}
\Delta \widetilde{E}[\Phi_{\rm TK}]= -\frac{\hbar m}{8\pi} \lim_{{\rm N_t}\to +\infty}\sum_{n=2}^{\rm N_t} \sum_{a=1}^N c_n^a(\mathbb{K})\, v_a^{2(1-n)}\, \Gamma[n-1] - \frac{\hbar m}{\pi} \sum_{\ell=1}^{N_{zm}} \sum_{a=1}^N v_a \,f_\ell^a(\mathbb{K}) \, \, ,
\label{final}
\end{equation}
where $N_t$ is the truncation order.

\subsection{One-loop $\Phi_{\rm TK}(\overline{x};\gamma)$-topological kink mass corrections}

In this Section we shall apply formula (\ref{final}) to compute the one-loop $\Phi_{\rm TK}(\overline{x};\gamma)$-kink mass shifts. We collect the needed data:

\vspace{0.2cm}

\noindent (1) The particle mass matrix  for the fluctuations around the $\Phi^{c(1)}$/$\Phi^{c(2)}$ vacua is:
\[
\mathbf{v}^2 = \left( \begin{array}{cc} 4& 0 \\ 0 & \sigma^2  \end{array} \right)\hspace{0.5cm};\hspace{0.5cm} v_1=2\hspace{0.5cm}, \hspace{0.5cm} v_2=\sigma \hspace{0.4cm}.
\]

\vspace{0.2cm}

\noindent (2) The $\Phi_{\rm TK}(\overline{x};\gamma)$-kink profile:
\[
\Phi_{\rm TK}(\overline{x};\gamma)=\left(\begin{array}{c} \widetilde{\phi}_1(\overline{x};\gamma)\\ \widetilde{\phi}_2(\overline{x};\gamma)\end{array}\right) \, \, .
\]
Although the kink orbits are  known explicitly, see (\ref{orbitkink}), there are no explicit analytic formulas for every topological kink profile in the $\gamma$-family except if $\sigma=2$ or $\sigma=\frac{1}{2}$, see \cite{Alonso2004b}. In the generic case, however, we may identify the BPS kink profiles by numerically solving the first-order differential equations (\ref{bps2}) with the initial condition $\Phi_{\rm TK}(0;\gamma)={0 \choose \frac{\gamma}{\sqrt{2\sigma}}}$ for any value of $\gamma\in[0,1)$.

\vspace{0.2cm}

\noindent (3) We thus write the $\mathbb{K}[\Phi_{\rm TK}(\overline{x};\gamma)]$-kink fluctuation operators (\ref{operatorK}) in terms of the numerically generated profiles. In this case the matrix potential wells are
\[
\mathbf{V}(\overline{x};\gamma) = \left( \begin{array}{cc}  -6+24\widetilde{\phi}_1^2(\overline{x};\gamma) +4\sigma(\sigma+1)\widetilde{\phi}_2^2(\overline{x};\gamma) & 8\sigma(\sigma+1)\widetilde{\phi}_1 (\overline{x};\gamma)\widetilde{\phi}_2(\overline{x};\gamma) \\ 8\sigma(\sigma+1)\widetilde{\phi}_1(\overline{x};\gamma) \widetilde{\phi}_2(\overline{x};\gamma) &  \sigma(1+\sigma)[4\widetilde{\phi}_1^2(\overline{x};\gamma)-1 ] +6\sigma^2 \widetilde{\phi}_2^2(\overline{x};\gamma) \end{array} \right)
\]

\noindent (4) The zero mode wave functions satisfy the linear ODE system (\ref{edozero}), i.e.,
\begin{equation}
A\,\Xi_{0\ell}=0 \quad \equiv \quad
    \left( \begin{array}{cc} -\frac{d}{dx} +4 \widetilde{\phi}_1(\overline{x};\gamma) & 2\sigma \widetilde{\phi}_2(\overline{x};\gamma) \\ 2\sigma \widetilde{\phi}_2(\overline{x};\gamma) & -\frac{d}{dx} +2 \sigma \widetilde{\phi}_1(\overline{x};\gamma) \end{array} \right) \left( \begin{array}{c} \xi_{0\ell}^1(\overline{x};\gamma) \\ \xi_{0\ell}^2(\overline{x};\gamma) \end{array} \right) = \left( \begin{array}{c} 0 \\ 0 \end{array} \right)\, \, . \label{sedo}
    \end{equation}
The numerical solutions of (\ref{sedo}) with the pair of initial conditions (a) $\xi_{01}^1(0;\gamma)=1$, $\xi_{01}^2(0;\gamma)=0$ and (b) $\xi_{02}^1(0;\gamma)=0$, $\xi_{02}^2(0;\gamma)=1$ form a set of two mutually orthogonal zero modes that we shall normalize properly.

\vspace{0.2cm}

These data are all what we need in the recurrence relations (\ref{recurrence2}) to generate the Seeley coefficients $c_n^a(\mathbb{K})$ and the coefficients $f_\ell^a(\mathbb{K})$ for every $\Phi_{\rm TK}(\overline{x};\gamma)$-kink in the $\gamma$-family. The one-loop shifts in the energy of any BPS kink in this family are finally estimated by means of the formula (\ref{final}) for a certain truncation order $N_t$. For instance, in Table 1 we display the kink shifts up to the truncation order $N_t=9$ for thirteen values of the parameter $\gamma\in (0,1)$ and eighteen values of the coupling constant $\sigma \in [1.4,3.1]$.

\begin{table}[ht]
\centerline{\small \begin{tabular}{|c|c|c|c|c|c|c|} \hline
\multicolumn{7}{|c|}{$\Delta E[\Phi_{\rm TK}(\overline{x};\gamma)]$} \\ \hline
$\gamma$ & $\sigma=1.4$ & $\sigma=1.5$ & $\sigma=1.6$ & $\sigma=1.7$ & $\sigma=1.8$ & $\sigma=1.9$ \\ \hline
$0.01$   & $-1.11623$ & $-1.15060$ & $-1.18560$ & $-1.22130$ & $-1.25768$ & $-1.29474$\\ \hline
$0.1$    & $-1.11687$ & $-1.15110$ & $-1.18598$ & $-1.22157$ & $-1.25785$ & $-1.29482$\\ \hline
$0.2$    & $-1.11881$ & $-1.15255$ & $-1.18712$ & $-1.22238$ & $-1.25836$ & $-1.29506$\\ \hline
$0.3$    & $-1.12203$ & $-1.15512$ & $-1.18901$ & $-1.22372$ & $-1.25920$ & $-1.29546$\\ \hline
$0.4$    & $-1.12655$ & $-1.15856$ & $-1.19164$ & $-1.22556$ & $-1.26035$ & $-1.29599$\\ \hline
$0.5$    & $-1.13241$ & $-1.16315$ & $-1.19500$ & $-1.22789$ & $-1.26178$ & $-1.29665$\\ \hline
$0.6$    & $-1.13969$ & $-1.16870$ & $-1.19905$ & $-1.23067$ & $-1.26346$ & $-1.29742$\\ \hline
$0.7$    & $-1.14861$ & $-1.17531$ & $-1.20385$ & $-1.23388$ & $-1.26536$ & $-1.29825$\\ \hline
$0.8$    & $-1.15978$ & $-1.18356$ & $-1.20956$ & $-1.23757$ & $-1.26745$ & $-1.29912$\\ \hline
$0.9$    & $-1.17525$ & $-1.19472$ & $-1.21705$ & $-1.24215$ & $-1.26985$ & $-1.30001$\\ \hline
$0.99$   & $-1.21768$ & $-1.22320$ & $-1.23497$ & $-1.25217$ & $-1.27439$ & $-1.30128$\\ \hline
$0.999$  & $-1.25723$ & $-1.24974$ & $-1.25142$ & $-1.26115$ & $-1.27828$ & $-1.30222$\\ \hline
$0.9999$ & $-1.29659$ & $-1.27615$ & $-1.26778$ & $-1.27008$ & $-1.28213$ & $-1.30315$\\ \hline
\end{tabular}}

\vspace{0.3cm}

\centerline{\small\begin{tabular}{|c|c|c|c|c|c|c|} \hline
\multicolumn{7}{|c|}{$\Delta E[\Phi_{\rm TK}(\overline{x};\gamma)]$} \\ \hline
$\gamma$ & $\sigma=2.0$ & $\sigma=2.1$ & $\sigma=2.2$ & $\sigma=2.3$ & $\sigma=2.4$ & $\sigma=2.5$ \\ \hline
$0.01$   & $-1.33251$ & $-1.37099$ & $-1.41020$ & $-1.45014$ & $-1.49081$ & $-1.53224$\\ \hline
$0.1$    & $-1.33251$ & $-1.37092$ & $-1.41006$ & $-1.44994$ & $-1.49057$ & $-1.53195$\\ \hline
$0.2$    & $-1.33251$ & $-1.37070$ & $-1.40966$ & $-1.44937$ & $-1.48986$ & $-1.53112$\\ \hline
$0.3$    & $-1.33251$ & $-1.37036$ & $-1.40901$ & $-1.44846$ & $-1.48872$ & $-1.52980$\\ \hline
$0.4$    & $-1.33251$ & $-1.36989$ & $-1.40815$ & $-1.44727$ & $-1.48726$ & $-1.52811$\\ \hline
$0.5$    & $-1.33251$ & $-1.36934$ & $-1.40713$ & $-1.44589$ & $-1.48560$ & $-1.52626$\\ \hline
$0.6$    & $-1.33251$ & $-1.36872$ & $-1.40605$ & $-1.44446$ & $-1.48395$ & $-1.52451$\\ \hline
$0.7$    & $-1.33251$ & $-1.36810$ & $-1.40501$ & $-1.44319$ & $-1.48262$ & $-1.52329$\\ \hline
$0.8$    & $-1.33251$ & $-1.36756$ & $-1.40421$ & $-1.44242$ & $-1.48216$ & $-1.52337$\\ \hline
$0.9$    & $-1.33251$ & $-1.36723$ & $-1.40409$ & $-1.44301$ & $-1.48390$ & $-1.52672$\\ \hline
$0.99$   & $-1.33251$ & $-1.36783$ & $-1.40702$ & $-1.44988$ & $-1.49624$ & $-1.54595$\\ \hline
$0.999$  & $-1.33251$ & $-1.36872$ & $-1.41051$ & $-1.45757$ & $-1.50962$ & $-1.56644$\\ \hline
$0.9999$ & $-1.33251$ & $-1.36961$ & $-1.41400$ & $-1.46525$ & $-1.52300$ & $-1.58692$\\ \hline
\end{tabular}}

\vspace{0.3cm}

\centerline{\small\begin{tabular}{|c|c|c|c|c|c|c|} \hline
\multicolumn{7}{|c|}{$\Delta E[\Phi_{\rm TK}(\overline{x};\gamma)]$} \\ \hline
$\gamma$ & $\sigma=2.6$ & $\sigma=2.7$ & $\sigma=2.8$ & $\sigma=2.9$ & $\sigma=3.0$ & $\sigma=3.1$ \\ \hline
$0.01$   & $-1.57442$ & $-1.61735$ & $-1.66105$ & $-1.70551$ & $-1.75074$ & $-1.79675$\\ \hline
$0.1$    & $-1.57410$ & $-1.61700$ & $-1.66067$ & $-1.70512$ & $-1.75033$ & $-1.79633$\\ \hline
$0.2$    & $-1.57315$ & $-1.61597$ & $-1.65958$ & $-1.70397$ & $-1.74915$ & $-1.79513$\\ \hline
$0.3$    & $-1.57168$ & $-1.61438$ & $-1.65790$ & $-1.70223$ & $-1.74739$ & $-1.79336$\\ \hline
$0.4$    & $-1.56984$ & $-1.61243$ & $-1.65588$ & $-1.70020$ & $-1.74538$ & $-1.79141$\\ \hline
$0.5$    & $-1.56787$ & $-1.61041$ & $-1.65389$ & $-1.69830$ & $-1.74364$ & $-1.78990$\\ \hline
$0.6$    & $-1.56612$ & $-1.60878$ & $-1.65248$ & $-1.69721$ & $-1.74295$ & $-1.78970$\\ \hline
$0.7$    & $-1.56517$ & $-1.60825$ & $-1.65251$ & $-1.69794$ & $-1.74452$ & $-1.79224$\\ \hline
$0.8$    & $-1.56604$ & $-1.61013$ & $-1.65562$ & $-1.70247$ & $-1.75068$ & $-1.80021$\\ \hline
$0.9$    & $-1.57139$ & $-1.61788$ & $-1.66615$ & $-1.71613$ & $-1.76781$ & $-1.82114$\\ \hline
$0.99$   & $-1.59887$ & $-1.65489$ & $-1.71390$ & $-1.77580$ & $-1.84050$ & $-1.90794$\\ \hline
$0.999$  & $-1.62781$ & $-1.69354$ & $-1.76348$ & $-1.83746$ & $-1.91536$ & $-1.99703$\\ \hline
$0.9999$ & $-1.65673$ & $-1.73218$ & $-1.81303$ & $-1.89909$ & $-1.99015$ & $-2.08606$\\ \hline
\end{tabular}}
\caption{\small One-loop $\Phi_{\rm TK}(\overline{x};\gamma)$-kink mass shifts for several values of the family parameter $\gamma$ in the coupling constant range $\sigma\in [1.4,3.1]$.}
\end{table}

\vspace{0.2cm}

The kink mass shift values shown in Table 1 plus the incoming graphics extracted from the Table are of great help in the qualitative interpretation of our results:

\vspace{0.2cm}

\noindent (1) The graphic of the simple $\Phi_{\rm TK}(x;0)$-kink one-loop shift as a function of $\sigma$ is shown in Figure 4. The red solid line represents the one-loop mass shift obtained in \cite{Alonso2004b} by means of the DHN formula. The modified GDW expansion estimations of the $\Phi_{\rm TK}(x;0.01)$-kink mass shifts are depicted, also $\sigma$-dependent, as open blue dots from the numbers in the first row of Table 1. The precision attained using the modified heat trace for the $\Phi_{\rm TK}(x;0.01)$-kink and tested against the exact result for the very close BPS kink $\Phi_{\rm TK}(x;0)$-kink in the moduli space is reassuring. Therefore, one expects that the modified asymptotic procedure opens the possibility of achieving reliable estimations of the kink mass corrections for the rest of the members of the $\Phi_{\rm TK}(\overline{x};\gamma)$-kink family, where no results are accessible by application of the exact DHN formula.

\begin{figure}[h]
\centerline{\includegraphics[width=6cm]{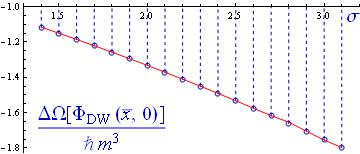}}
\caption{Comparison between the $\Phi_{\rm TK} (\overline{x};0)$-kink mass quantum correction computed by means of the DHN formula, see \cite{Alonso2004b}, (solid line) and the asymptotic value of the $\Phi_{\rm TK} (\overline{x};0.01)$-kink mass shift afforded by zeta function methods adapted to the existence of zero modes (blue dots).}
\end{figure}

\vspace{0.2cm}

\noindent (2) In Figure 5 we show the variability of the one-loop $\Phi_{\rm TK}(\overline{x};\gamma)$-kink mass shift
for different $\gamma$'s as function of $\sigma$ in the range $\gamma\in[0,0.9999]$. The classical energy degeneracy between all the $\Phi_{\rm TK}(\overline{x};\gamma)$-kink family members is broken at the one-loop level for almost any value of $\sigma$, except for the particular case $\sigma=2$. For $\sigma\neq 2$, the one-loop mass correction depends on $\gamma$. It appears that a quantum phase transition is induced by the kink one-loop fluctuations breaking the classical degeneracy in kink mass.

\begin{figure}[h]
\centerline{\hspace{1cm}
\includegraphics[width=5cm]{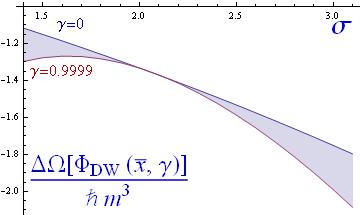}}
\caption{Variability in $\gamma$ of the kink mass quantum shift as function of $\sigma\in [1.4,3.1]$.}
\end{figure}

\noindent (3)  The surviving degeneracy at $\sigma=2$ suggests that the $\sigma <2$ and $\sigma>2$ regimes should be analyzed in turn:

\begin{itemize}
    \item $\sigma<2$: In Figure 6(a) ($\sigma=1.5$) we see that the shifts in the kink mass decrease from the simple kink $\Phi_{\rm TK}(x;0)$ energy with increasing values of $\gamma$. The quantum fluctuations induce an outwards force between the two components of the BPS kink. A repulsive Casimir force arises between the two basic lumps when the kink mass diminishes towards the $\gamma\to 1$ kink.{\footnote{$\gamma=1$ cannot be reached because this would imply a change of topological sector, requiring infinite energy, or, from another perspective, the basic lumps are infinitely separated, a process forbidden by the topology of the configuration space.}}

\begin{figure}[h]
\centerline{\includegraphics[width=3.6cm]{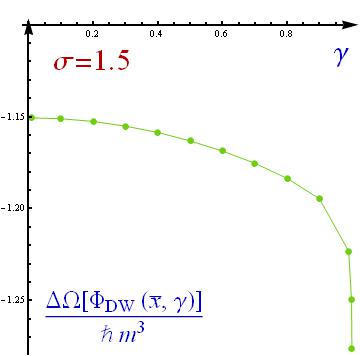}\hspace{1cm}
\includegraphics[width=3.6cm]{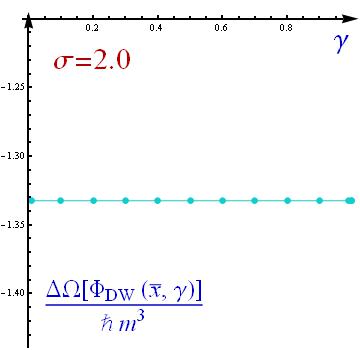}\hspace{1cm}
\includegraphics[width=3.6cm]{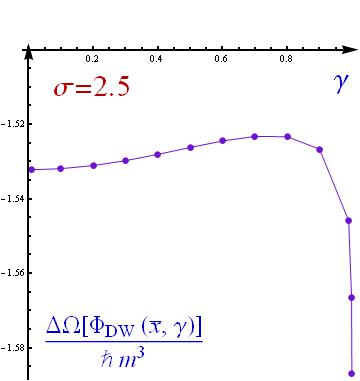}}
\caption{One-loop mass shifts for the cases (a) $\sigma=1.5$, (b) $\sigma=2.0$ and (c) $\sigma=2.5$.}
\end{figure}

    \item If the coupling constant is such that $\sigma>2$ the situation is more sophisticated, see Figure 6(c) ($\sigma=2.5$). If the $\Phi_{\rm TK}(\overline{x};\gamma)$-kink is a one-lump configuration, i.e., $\gamma$ is small, the quantum fluctuations induce an inwards force between the components because the kink mass augments when $\gamma$ increases.  An attractive Casimir force arises that tends to merge the constituent lumps into the $\Phi_{\rm TK}(\overline{x};0)$ simple kink where the two lumps fully overlap. It is clear in Figure 6(c) that at $\Phi_{\rm TK}(\overline{x};0)$ a local minimum of the wall tension is reached; this simplest kink is lighter than the other walls in its neighborhood. Things are different, however, for larger values of $\gamma$. There exists a critical distance fixed by a critical $\gamma$ such that if the two basic lumps are initially more separated than this distance, the quantum fluctuations induce a force outwards between them and push the two extended objects apart farther.

\item At the critical value $\sigma=2$ the kink mass classical degeneracy is preserved at the one loop level, see Figure 6(b). Thus, $\sigma=2$ is a critical point characterizing a very peculiar phase transition: the attractive forces between the two extended lumps when $\sigma>2$ turn into repulsive forces which tend to separate these lumps for $\sigma<2$, at least if the composite kink is a configuration formed by two not too distant lumps. The phase transition thus presents clear similarities with the transition from Type I ($\gamma\ll 1$, $\sigma>2$) to Type II
($\sigma<2$) superconductors. There are three differences: (1) In this case the phase transition is of quantum mechanical nature. (2) Two magnetic flux lines in Type I superconductivity always attract each other, regardless the distance between them. If $\sigma>2$ the basic kinks in this model attract each other if the relative distance is small but there is repulsion beyond a critical distance between their center of masses. (3) In this model, the count of basic extended objects ends in two.

We finish this Section by explaining an analytical peculiarity arising at $\sigma=2$, responsible for the survival of the kink mass degeneracy. The potential energy density (\ref{bnrtpotential}) for this particular case $U(\phi_1,\phi_2)= \frac{1}{8} (4\phi_1^2+4 \phi_2^2-1)^2+8 \phi_1^2 \phi_2^2$ decouples after performing a $\frac{\pi}{4}$-rotation in field space: $\varphi_1=\frac{1}{\sqrt{2}}(\phi_1+\phi_2)$ and $\varphi_2=\frac{1}{\sqrt{2}}(\phi_1-\phi_2)$. In the new variables $\varphi_1$ and $\varphi_2$ the potential energy density reads $U(\varphi_1,\varphi_2)= 4 ( \varphi_1^2-\frac{1}{8} ) + 4 ( \varphi_2^2-\frac{1}{8})$, whereas the BPS kink family becomes:
\[
\Upsilon_{\rm TK}(\overline{x};\gamma)=\left(\begin{array}{c} \varphi_1(x) \\ \varphi_2(x)\end{array}\right)= \frac{1}{2\sqrt{2}}\Big(\begin{array}{c}  \tanh \overline{x} \\ \tanh (\overline{x}+\gamma)\end{array}\Big) \, \, .
\]
$\gamma$ is not exactly the same parameter as before but plays the same r$\hat{\rm o}$le characterizing the different kinks in the family. The $\Upsilon_{\rm TK}(\overline{x};\gamma)$-kink fluctuation operator
\begin{equation}
\mathbb{K}[\Upsilon_{\rm TK}(\overline{x};\gamma)]=\left( \begin{array}{cc} -\frac{d^2}{dx^2}+4-6\, {\rm sech}^2 \overline{x} & 0 \\ 0 & -\frac{d^2}{dx^2}+4 -6 \, {\rm sech}^2 (\overline{x} + \gamma)\end{array} \right) \label{tk2hess}
\end{equation}
is diagonal for all $\gamma$!!. Moreover, the DNH formula can be applied to (\ref{tk2hess}) in order to exactly compute the quantum corrections to all kink masses. Both differential operators (\ref{tk2hess}) on the diagonal are also the second member in the hierarchy (two bound states) of transparent P$\ddot{\rm o}$schl-Teller operators for any $\gamma$!!. Thus, we easily obtain the $\gamma$-independent $\Delta \widetilde{E}[\Upsilon_{\rm TK}(\overline{x};\gamma)]=-1.33251 \hbar m$ kink mass shift from the DHN formula. We stress that this exact result coincides with the estimation displayed in Table 1 derived from the modified GDW expansion.
\end{itemize}

\section{One-loop surface tension shifts of classical BPS domain walls}

We shall study now the model in
$\mathbb{R}^{1,3}$ Minkowski space-time where the BPS solutions of Section \S.2 are kink domain walls, two-dimensional extended objects grown from the kink point defects through symmetry in the $x^2$ and $x^3$ directions. The domain wall fluctuations, respectively the vacuum fluctuations, are governed by the partial differential operator $\mathbb{L}$, respectively  $\mathbb{L}_0$, defined in Section \S.2.1. In particular, the second goal in this work is the evaluation of the surface tension corrections due to one-loop quantum fluctuations of the classically degenerate domain walls arising in the $N=2$ scalar field model described in sub-Section \S.2.2. To achieve this task we shall develop the heat kernel/zeta function approach adapted to the matrix partial differential operators $\mathbb{L}$ and $\mathbb{L}_0$ obtained by enlarging the kink and vacuum fluctuations $\mathbb{K}$ and $\mathbb{K}_0$ to the Euclidean space $\mathbb{R}^3$.
The second-order differential operator governing the fluctuations around the $\Phi_{\rm DW}(x^1;0)$ domain wall is diagonal:
\begin{equation}
\mathbb{L}[\Phi_{\rm DW}(x^1;0)]=\left( \begin{array}{cc} -\nabla^2+4-6 \, {\rm sech}^2 x^1 & 0 \\ 0 & -\nabla^2+\sigma^2-\sigma(\sigma+1) \, {\rm sech}^2 x^1  \end{array} \right)\hspace{0.4cm}.
\label{tk1operator}
\end{equation}
It is clear that both the heat trace and zeta functions of $\mathbb{L}_0$ are easily given in terms of the heat trace and zeta functions of $\mathbb{K}_0$:
\begin{equation*}
h_{\mathbb{L}_0}(s)=\frac{l^3}{8 \pi^\frac{3}{2}} \frac{1}{\beta^{\frac{3}{2}}}\sum_{a=1}^N e^{-\beta v_a^2}=\frac{l^2}{4\pi\beta}h_{\mathbb{K}_0}(\beta) \quad , \quad \zeta_{\mathbb{L}_0}(s)=\frac{l^2}{4\pi}\frac{\Gamma(s-1)}{\Gamma(s)}\zeta_{\mathbb{K}_0}(s-1) \quad .
\end{equation*}
Analogous relationships exist between the heat traces and zeta functions of the domain wall and kink fluctuation operators because there are no dependence on $x^2$ and $x^3$ in $\mathbf{V}(\bar{x}^1)$ and the parallel to the wall eigenfunctions are plane waves:
\begin{eqnarray*}
h_{\mathbb{L}}(s)&=&\frac{l^2}{4\pi}\int_{-\infty}^\infty dk_2 \int_{-\infty}^\infty  \frac{l^2}{(2\pi)^2} e^{-\beta(k_2^2+k_3^2)} h_{\mathbb{K}}(\beta)=\frac{l^2}{4\pi\beta}h_{\mathbb{K}}(\beta) \hspace{0.4cm}, \\ \zeta_{\mathbb{L}}(s)&=&\frac{l^2}{4\pi}\frac{\Gamma(s-1)}{\Gamma(s)}\zeta_{\mathbb{K}}(s-1) \quad .
\end{eqnarray*}
Thus, we shall profit from the previous results on heat traces and zeta functions for the kink fluctuation operators $\mathbb{K}$ in the application of the zeta function method to the estimation of domain wall surface tension shifts.

\subsection{One-loop surface tension shift of the simplest kink domain wall $\mathbf{\Phi}_{\rm DW}(\bar{x}^1;0)$}

In this particular case the domain wall fluctuation operators is:
\[
\mathbb{L}[\mathbf{\Phi}_{\rm DW}(\bar{x}^1;0)]= \left(\begin{array}{cc} \mathbb{L}_1 & 0 \\ 0 & \mathbb{L}_2\end{array}\right) =\left(\begin{array}{cc} -\nabla^2+4 -6{\rm sech}^2 \bar{x}^1 & 0 \\ 0 & -\nabla^2+\sigma^2-\sigma(\sigma+1){\rm sech}^2\bar{x}^1\end{array}\right) \, \, .
\]
Given the diagonal structure as a matrix we only need to deal with the spectra of differential operators of the type:
\begin{equation*}
\mathbb{L}_a=-\nabla^2+v_a^2-v_a(v_a+1) \, {\rm sech}^2\bar{x}^1 \, \, , \, \, a=1,2 \quad , \quad v_1=2 \, \, , \, \, v_2=\sigma
\end{equation*}
to calculate the associated spectral functions. We shall
perform the analysis when $\sigma=J\in\mathbb{N}^*$ is a non-zero natural number. The reason is that in this case all the spectral data are given in terms of well known special functions and it is possible to obtain analytical formulas. Notice also that $\mathbb{L}_1$ is the $v_1=J=2$ case in the hierarchy of reflectionless P$\ddot{\rm o}$sch-Teller Hamiltonians. The spectrum of the $\mathbb{L}_a(\sigma=J)$ operator is summarized as follows:

\vspace{0.2cm}

\noindent (1) Drifted zero mode: $\omega^2_2(i J,k_2,k_3)=k_2^2+k_3^2$ and
$f_{\small{i J,k_2,k_3}}(x^1,x^2,x^3)= {\rm sech}^Jx^1 e^{i(k_2x^2+k_3x^3)}$ are the eigenvalues and eigenfunctions describing the propagation in the $x^2:x^3$ parallel plane of the wall translational mode along the $x^1$-axis .
\vspace{0.2cm}

\noindent (2) Drifted bound states: $\omega_2^2(i (J-j),k_2,k_3)=(2J-j)j +k_2^2+k_3^2$, $j=1,2, \cdots, J-1$, and
\begin{eqnarray*}
&& f_{\small{i(J-j),k_2,k_3}}(x^1,x^2,x^3)=e^{i(k_2x^2+k_3x^3)}\times \left(-\frac{\partial}{\partial x^1}+(J-1){\rm
tanh}\,x^1\right) \times \\ && \hspace{1cm} \times\left(-\frac{\partial}{\partial x^1}+(J-2){\rm
tanh}\,x^1\right)\cdots \left(-\frac{\partial}{\partial x^1}+{\rm tanh}\,x^1\right){\rm sech}^{J-j}x^1
\end{eqnarray*}
are the eigenvalues and eigenfunctions corresponding to the propagation of the positive energy bound states -mesons trapped by the one-dimensional kink well- in the parallel directions to the wall.
\vspace{0.2cm}

\noindent (3) Mesons crossing orthogonally the wall are described by scattering waves. The drift parallel to the wall of these waves give rise  to the eigen-functions:
\begin{eqnarray*}
&& f_{\small{k_1,k_2,k_3}}(x^1,x^2,x^3)=e^{i(k_2x^2+k_3x^3)} \times \left(-\frac{\partial}{\partial x^1}+J \, {\rm
tanh} \, x^1\right) \times   \left(-\frac{\partial}{\partial x^1}+(J-1){\rm
tanh} \, x^1\right) \\ && \hspace{3cm} \times \cdots \times \left(-\frac{\partial}{\partial x^1}+{\rm tanh} \, x^1 \right)e^{ik_1x^1} = P_J({\rm tanh} \, x^1 ,k_1)e^{i(k_1x^1+k_2x^2+k_3x^3)}\, \, ,
\end{eqnarray*}
with $\omega_2^2(k_1,k_2,k_3)=k_1^2 +k_2^2+k_3^2 +J^2$. From  $P_J({\rm tanh}x^1,k_1)$ - the $J$-th Jacobi polynomial times- we read both the scattering phase shift produced by the wall and the spectral density:
\[
\delta(k_1)=2{\rm arctan}\left[\frac{{\rm
Im}(\Pi_{j=0}^{J-1}(J-j-ik_1))}{{\rm Re}(\Pi_{j=0}^{J-1}(J-j-ik_1))}\right] \hspace{0.2cm},\hspace{0.2cm}
\rho(k_1)= \frac{l}{2\pi}-\frac{1}{\pi}\sum_{j=0}^{J-1}
\frac{J-j}{(J-j)^2+k_1^2} \, .
\]
We recall that the scattering is transparent such that the phase shifts in the even and odd channels are equal, see \cite{Alonso2012b}.

\vspace{0.2cm}

\noindent With all this information we calculate:
\[
h_{\mathbb{L}_2(J)}(\beta)-h_{\mathbb{L}_{02}}(\beta)= \frac{l^2}{4\pi\beta}e^{-\beta J^2}\sum_{j=0}^{J-1}\, e^{\beta(J-j)^2}{\rm Erf}[(J-j)\sqrt{\beta}] \label{hfdw2} \, \, .
\]
It is clear from the integral in the first row of this formula that the effect of the parallel fluctuations to the wall is not completely balanced by the vacuum fluctuations because the spectral densities differ due to the phase shifts acquired by the meson waves in the orthogonal crossing of the domain wall. Mellin transform of formula (\ref{hfdw2}) provides us with the domain wall zeta function:
\begin{eqnarray}
\zeta_{\mathbb{L}_2(J)}(s)-\zeta_{\mathbb{L}_{02}}(s) &=& \frac{l^2}{4\pi^\frac{3}{2}} \frac{\Gamma(s-\frac{1}{2})}{\Gamma(s)}\times \Big\{-\frac{1}{J^{2s-2}(s-1)} +
\label{zfdw}\\ &+& \sum_{j=1}^{J-1}\frac{2(J-j)}{((2J-j)j)^{s-\frac{1}{2}}} {}_2F_1[ {\textstyle\frac{1}{2},s -\frac{1}{2},\frac{3}{2}, -\frac{(J-j)^2}{(2J-j)j} }] \Big\} \nonumber \, .
\end{eqnarray}
The Casimir domain wall energy per unit of surface due to fluctuations in the field space $\phi_2$-direction is subsequently regularized in the framework of the zeta function method as:
\begin{eqnarray*}
\Delta\widetilde{\Omega}^{(2)}_1[\mathbf{\Phi}_{\rm DW}(\bar{x}^1,0)](s)=\frac{\hbar m}{2L^2}\left(\frac{\mu^2}{m^2}\right)^{s+\frac{1}{2}}\left(\zeta_{\mathbb{L}_2(J)}(s)-\zeta_{\mathbb{L}_{02}}(s)\right) \, \, .
\end{eqnarray*}
In order to isolate the divergences arising at the physical point $s=-\frac{1}{2}$, which is a pole of $\Delta\widetilde{\Omega}^{(2)}_1[\mathbf{\Phi}_{\rm DW}(\bar{x}^1,0)](s)$, we estimate this magnitude in a point of the $s$-complex plane in the neighborhood of this pole:
\begin{eqnarray}
&& \Delta\widetilde{\Omega}^{(2)}_1[\mathbf{\Phi}_{\rm DW}(\bar{x}^1,0)](-{\textstyle\frac{1}{2}}+\varepsilon)=\frac{\hbar m^3}{8 \pi^\frac{3}{2}}\cdot \Big\{ \frac{2 J^4+J^3-J^2}{6}\cdot\frac{1}{\varepsilon}+\frac{J^3}{3}\Big(\frac{7}{6}-\log J^2\Big) - \nonumber  \\ && -\sum_{j=1}^{J-1}\, j(J-j) (2J-j)\Big[\frac{2j^2-4jJ-J^2}{3j(j-2J)}\log[2J-j] -  {}_2F_1^{(0,1,0,0)}[{\textstyle\frac{1}{2},-1,\frac{3}{2},- \frac{(J-j)^2}{(2J-j)j}}]\Big]+ \nonumber \\ &&+ \frac{2 J^4+J^3-J^2}{6} \Big(\log {\textstyle \frac{\mu^2}{m^2}}-\gamma_E-\psi({\textstyle -\frac{1}{2}})\Big) +\frac{J^2}{3}(J^2-{\textstyle\frac{1}{2}})+{\cal O}(\varepsilon)\Big\}\, \, .\label{uvdcdw}
\end{eqnarray}
The total Casimir $\mathbf{\Phi}_{\rm DW}(x^1;0)$-domain wall surface tension is obtained by adding the effect of the $\phi_1$-fluctuations
accounted for by the formula above in the $J=2$ case. The next task is to choose a renormalization criterion to tame the ultraviolet divergences shown in the first row of formula (\ref{uvdcdw}). Guided by the experience in $(1+1)$-dimensions we choose the heat kernel renormalization criterion, see  \cite{Bordag2002}.

The modified Gilkey-de Witt heat kernel expansion also works in $(3+1)$-dimensions for backgrounds depending only on one coordinate. Besides the shift in the powers of $\beta$ due to the jump in dimensions one needs to compute the Seeley coefficients -after solving the modified recurrence relations- via a volume integral. Because the symmetry of the domain walls the new coefficients are merely the old ones times the normalizing area $l^2$. Thus, we write the Casimir wall surface tension -the Casimir energy per unit of area- alternatively in the form
\begin{eqnarray}
&& \Delta \widetilde{\Omega}_1[\Phi_{DW}](s) =  \frac{\hbar m^3}{2}\Big( \frac{\mu^2}{m^2} \Big)^{s+\frac{1}{2}} \frac{1}{8\pi^{\frac{3}{2}}}  \Big[  \sum_{a=1}^2 c_1^a(\mathbb{K}) (v_a^2)^{\frac{1}{2}-s} \frac{\Gamma[s-\frac{1}{2}]}{\Gamma[s]} +\nonumber\\ && \hspace{1cm} + \sum_{a=1}^2 c_2^a(\mathbb{K}) (v_a^2)^{-\frac{1}{2}-s} \frac{\Gamma[s+\frac{1}{2}]}{\Gamma[s]} + \sum_{n=3}^\infty \sum_{a=1}^2 c_n^a(\mathbb{K}) (v_a^2)^{\frac{3}{2}-n-s} \frac{\Gamma[s+n-\frac{3}{2}]}{\Gamma[s]} +\nonumber \\ && \hspace{1cm} + \sum_{\ell=1}^{N_{zm}} \sum_{a=1}^2 f_\ell^a(\mathbb{K}) (v_a^2)^{1-s} \frac{\Gamma[s-\frac{1}{2}]}{(1-s)\Gamma[s]} \Big] \hspace{1cm} . \label{dwcmhk}
\end{eqnarray}
The powers of $v_a$ multiplying respectively the first and second Seeley coefficients are $2$ and $0$. {\footnote{The argument does not include the contribution of the zero modes appearing in the heat kernel coefficients derived by means of the modified recurrence relations. Only the traditional first two Seeley coefficients must be canceled. The reason is that zero modes exist in the $m\to\infty$ range. Kinks become singular rigid Heaviside step and/or Dirac delta backgrounds but still have translational modes that leave an infinite contribution by themselves.}} But these parameters are the particle masses. In the limit of infinite mass there are no fluctuations and therefore there cannot be any shift in the surface tension. We shall subtract accordingly the contributions of the two first terms in the sum that would survive even for (in the absence of) fluctuations of infinitely heavy particles. The same criterion in $(1+1)$-dimension only requires to kill the contribution of the first coefficient because the powers of $v_a$ entering in the first two terms are in that case respectively $0$ and $-2$; only the contribution of the first term survive in the mass infinite limit. It is known that in $(1+1)$-dimensional models this criterion is tantamount to the tadpole vanishing condition in a minimal renormalization scheme. The cancelation of the contribution of the (a) $a_1(\mathbb{K}_a)$ and (b) $a_2(\mathbb{K}_a)$ coefficients as renormalization criterion is equivalent in our $(3+1)$-dimensional model to take into account the effect of the counter-terms subtracting: (a) the quadratic divergences of the one-loop tadpole and 1-vertex one-loop self-energy graphs (b) the sub-dominant logarithmic divergences of these graphs plus the (also logarithmic) divergences of one-loop self-energy diagrams with two vertices, and the tri-valent and four-valent vertex corrections at one-loop order. In sum, the regularized contribution to the surface tension shift due to the counter-terms is:
\begin{equation*}
\Delta\widetilde{\Omega}^{(a)}_2[\mathbf{\Phi}_{\rm DW}(\bar{x}^1;0)](s)=\frac{\hbar m^3}{16 \pi^{\frac{3}{2}} \Gamma(s)}\left(\frac{\mu^2}{m^2}\right)^{s+ \frac{1}{2}} \!\! \int_0^\infty \!\!\! d\beta \, \left[\langle V_{aa}\rangle\beta^\frac{3}{2}+ \left\langle {\textstyle\frac{1}{6} V_{aa}^{\prime\prime}-\frac{1}{2}\left(V_{aa}\right)^2}\right\rangle\right]e^{-v_a^2\beta}
\end{equation*}
For the simple domain wall at the stake we have $v_2=J$ and $V_{22}(x^1)=-J(J+1){\rm sech}^2 x^1$, such that:
{\small\begin{equation*}
\Delta\widetilde{\Omega}^{(2)}_2[\mathbf{\Phi}_{\rm DW}(\bar{x}^1;0)]=-\frac{\hbar m^3}{16 \pi^{\frac{3}{2}}}\left(\frac{\mu^2}{m^2}\right)^{s+\frac{1}{2}} \frac{1}{\Gamma(s)}\left[\frac{J(J+1)}{3 J^{2s-1}}\Gamma(s-{\textstyle \frac{1}{2}}) + \frac{2}{3}\frac{J^2(J+1)^2}{J^{2s+1}} \Gamma(s+{\textstyle\frac{1}{2}})\right] \, \, ,
\end{equation*}}
or, near the physical point,
\begin{eqnarray}
&& \hspace{-1cm} \Delta\widetilde{\Omega}^{(2)}_2[\mathbf{\Phi}_{\rm DW}(\bar{x}^1;0)]({\textstyle -\frac{1}{2}}+\varepsilon)= \nonumber \\
&& =-\frac{\hbar m^3}{16 \pi^{\frac{3}{2}}}\left(\frac{\mu^2}{m^2}\right)^{\varepsilon}\left[\frac{2J(J+1)}{J^{-2+2\varepsilon}}\frac{\Gamma(-1+\varepsilon)}{\Gamma(-\frac{1}{2}+\varepsilon)} + \frac{2}{3}\frac{J^2(J+1)^2}{J^{2\varepsilon}}\frac{\Gamma(\varepsilon)}{\Gamma(-\frac{1}{2}+\varepsilon)}\right] \nonumber \\ && \simeq -\frac{\hbar m^3}{8\pi^2} \left\{\frac{2 J^4+J^3-J^2}{6}\cdot \frac{1}{\varepsilon}-\frac{2 J^4+J^3-J^2}{6} \left[\psi(0,-{\textstyle\frac{1}{2}})- \log\left(\frac{\mu^2}{m^2}\right)\right] \right. \label{uvddw}
\\ && +\left. \frac{1}{4}\left[\frac{2}{3}J^2(J+1)^2(\gamma_E-\log J^2)-2 J^3(J+1)(-1+\gamma_E +\log J^2 )   \right]+ {\cal O}(\varepsilon)\right\}\nonumber \hspace{0.2cm} .
\end{eqnarray}
Both the divergences and the dependence in the $\mu$-parameter disappear in
\[
\Delta\widetilde{\Omega}^{(2)}[\mathbf{\Phi}_{\rm DW}(\bar{x}^1;0)](-{\textstyle\frac{1}{2}})=\Delta\widetilde{\Omega}^{(2)}_1[\mathbf{\Phi}_{\rm DW}(\bar{x}^1;0)](-{\textstyle\frac{1}{2}})+\Delta\widetilde{\Omega}^{(2)}_2[\mathbf{\Phi}_{\rm DW}(\bar{x}^1;0)](-{\textstyle\frac{1}{2}})
\]
leaving a finite remainder that gives the one-loop correction to the surface tension of the domain wall due to fluctuations of the simple domain wall in the $\phi_2$-axis in field space. In the case $J=2$, for instance, we obtain, adding the identical shift produced by fluctuations in the $\phi_1$ direction:
\[
\Delta\widetilde{\Omega}[\mathbf{\Phi}(\bar{x}^1;0)]\vert_{\sigma=2}=
\frac{\hbar m^3}{\pi^2}\left[-\frac{19}{18}+\frac{5}{6}\log\frac{4}{3}+\frac{3}{4}\cdot {}_2F_1^{(0,1,0,0)}[{\textstyle \frac{1}{2},-1,\frac{3}{2},-\frac{1}{3}}] \right]=-0.0918881 \hbar m^3
\]

\subsection{Quantum corrections of the surface tensions of generic BPS domain walls}

Finally, we offer the calculation of surface tension shifts of generic domain walls using the modifed heat kernel expansion derived in previous Sections. We shall write the pertinent formulas for any domain wall in a model with $N$ scalar fields to address later our $N=2$ model. The domain wall Casimir tension, the Casimir energy per unit of surface of the wall, is regularized in the spectral function framework
as:
\begin{equation*}
\Delta\widetilde{\Omega}_1[\mathbf{\Phi}_{DW}(\bar{x}^1)](s) = \frac{\hbar m}{2 L^2} \Big( \frac{\mu^2}{m^2} \Big)^{s+\frac{1}{2}} \left(\zeta_\mathbb{L}(s)-\zeta_{\mathbb{L}_0}(s)\right)
\end{equation*}
In the physical limit $s\rightarrow -\frac{1}{2}$ ultraviolet divergences arise. As explained in the previous sub-Section the control of these divergences will be achieved by means of the heat kernel renormalization criterion. Both in (1+1)- and (3+1)-dimensions is a minimal renormalization scheme but in the later case counter-terms must be introduced that tame, not only the tadpoles and self-energy graphs, but also the tri-valent and four-valent vertex corrections all of them at one-loop order. There are two contributions to the surface tension shifts due to renormalization counter-terms:

\vspace{0.2cm}

\noindent (1) The counter-terms that cancel the dominant divergences of tadpole and self-energy graphs give rise to the surface tension shift:
\begin{equation*}
\Delta \widetilde{\Omega}_2[\mathbf{\Phi}_{DW}(x^1)](s)=\frac{\hbar m^3}{2} \Big( \frac{\mu^2}{m^2} \Big)^{s+\frac{1}{2}} \sum_{a=1}^N \langle V_{aa}(x^1) \rangle \frac{(v_a^2)^{\frac{1}{2}-s}}{8 \pi^{\frac{3}{2}}} \frac{\Gamma[s-\frac{1}{2}]}{\Gamma[s]}
\end{equation*}
that kills the contribution of the $a_1^a(\mathbb{K})$ coefficients.

\vspace{0.2cm}

\noindent (2) The counter-terms that annihilate the divergences of the tri-valent and four-valent vertex corrections at one-loop plus the sub-dominant divergences of tadpoles and self-energy graphs cancel the contributions of the $a_2^a(\mathbb{K})$ coefficients:
\[
\Delta \widetilde{\Omega}_3[\mathbf{\Phi}_{DW}](s)=\frac{\hbar m^3}{16 \pi^{\frac{3}{2}}} \Big( \frac{\mu^2}{m^2 v_a^2} \Big)^{s+\frac{1}{2}} \sum_{a=1}^N \left\langle \Big({\textstyle \frac{1}{6} \mathbf{V}^{\prime\prime}(x^1)-\frac{1}{2} \mathbf{V}^2(x^1)-\frac{1}{2}[\mathbf{v}^2,\mathbf{V}(x^1)]}\Big)_{aa} \right\rangle  \frac{\Gamma[s+\frac{1}{2}]}{\Gamma[s]}
\]

Adding these two pieces of the surface tension corrections to the domain wall Casimir tension (\ref{dwcmhk}) we obtain the total shift
\[
\Delta{\widetilde{\Omega}}[\mathbf{\Phi}_{DW}] = \lim_{s\rightarrow -\frac{1}{2}} \Big[ \Delta \widetilde{\Omega}_1[\mathbf{\Phi}_{DW}](s) + \Delta\widetilde{\Omega}_2[\mathbf{\Phi}_{DW}](s)+ \Delta \widetilde{ \Omega}_3[\mathbf{\Phi}_{DW}](s) \Big]
\]
and having in mind the modified heat function expansion we finally write:
\begin{equation}
\Delta \widetilde{\Omega}[\Phi_{DW}]= - \frac{\hbar m^3}{18\pi^2}\sum_{\ell=1}^{N_{zm}} \sum_{i=1}^N f^a_\ell(\mathbb{K}) v_a^3 - \frac{\hbar m^3}{32\pi^2} \lim_{{\rm N_t}\rightarrow \infty} \sum_{n=3}^{{\rm N}_t} \sum_{a=1}^N c_n^a(\mathbb{K}) (v_a^2)^{2-n} \Gamma[n-2] \label{dwstsol}
\end{equation}
We feed this formula with the data obtained numerically for the BPS kinks in our $N=2$ model and used in the analogous formula for kinks (\ref{final}). The differences with (\ref{final}) are: (1) some different factors of $\pi$, $m$, etcetera, (2) the truncated series start at the third-order term, and (3) the arguments of the $\Gamma$ functions are shifted backwards one order. The results for the surface tension shifts of the classically degenerate BPS walls are collected in Table 2. A summary follows:

\vspace{0.2cm}

\noindent (1) As for BPS kinks the classical degeneracy is broken and  differences of pressures between basic walls at different distances are induced by  one-loop quantum fluctuations.

\vspace{0.2cm}

\noindent (2) The non-dimensional wall tension shifts are weaker as compared to the non-dimensional kink mass shifts by a factor between 10 and 20 in the range of the coupling constant that we have studied.

\vspace{0.2cm}

\noindent (3) Of course, the factor depends on the renormalization prescription but we find a qualitative result. If the coupling is weak the kink well
is weakly attractive. The parallel to the wall fluctuations are comparatively important. Thus, the shifts in surface tension are lower and the quotients are higher. When the coupling is strong, however, the transverse fluctuations to the wall dominate because the kink well is strongly attractive, the surface tension shifts grow and the quotients diminish. There is also a mild dependence of the quotients on $\gamma$: at fixed $\sigma$ the quotient between the kink mass and wall tension shifts is greater if $\gamma$ is close to $1$, i.e., if the two basic lumps are far apart.

\vspace{0.2cm}

\noindent (4) $\sigma=2$ is the only case where the classical degeneracy is not broken by quantum fluctuations. The reasons were explained at the end of Section \S.4 for kinks and work similarly in this three-dimensional setting for domain walls. Confidence in the heat kernel method is enhanced by the observation that the results given at the Table for the $\sigma=2$ surface tension shifts fit perfectly with the exact calculation of this magnitude for the $\gamma=0$-domain wall shown in the previous sub-Section.

\vspace{0.2cm}

\noindent (5) In Figure 7 it is shown that the induced pressure changes respond to attractive forces between the two basic walls up to values of $\gamma$ close to one (when the walls are far apart) both for $\sigma<2$ and $\sigma>2$. This behaviour is in contrast with the forces induced on kinks, see the corresponding Figure 6, where we find repulsion between the basic kinks whenever $\sigma<2$. There are hints of weak repulsion between domain walls for coupling constant values of the order of $\sigma=1.9$, just below the critical value $\sigma=2$, as it is possible to check at the Tables.

\begin{table}[ht]
\centerline{\small \begin{tabular}{|c|c|c|c|c|c|c|} \hline
\multicolumn{7}{|c|}{$\Delta \Omega[\mathbf{\Phi}_{\rm DW}(\overline{x}^1;\gamma)]$} \\ \hline
$\gamma$ & $\sigma=1.4$ & $\sigma=1.5$ & $\sigma=1.6$ & $\sigma=1.7$ & $\sigma=1.8$ & $\sigma=1.9$ \\ \hline
$0.01$   & $-0.0613943$ & $-0.0649785$ & $-0.0690983$ & $-0.0738001$ & $-0.0791321$ & $-0.0851443$\\ \hline
$0.1$    & $-0.0613599$ & $-0.0649578$ & $-0.069088$ & $-0.0737971$ & $-0.0791333$ & $-0.0851465$\\ \hline
$0.2$    & $-0.0612578$ & $-0.0648964$ & $-0.0690576$ & $-0.0737884$ & $-0.0791369$ & $-0.0851529$\\ \hline
$0.3$    & $-0.0610952$ & $-0.0647985$ & $-0.0690095$ & $-0.073775$ & $-0.0791431$ & $-0.0851686$\\ \hline
$0.4$    & $-0.0608829$ & $-0.0646711$ & $-0.0689473$ & $-0.0737584$ & $-0.0791521$ & $-0.0851781$\\ \hline
$0.5$    & $-0.0606358$ & $-0.0645233$ & $-0.0688761$ & $-0.0737405$ & $-0.079164$ & $-0.085196$\\ \hline
$0.6$    & $-0.0603736$ & $-0.0643672$ & $-0.0688021$ & $-0.0737238$ & $-0.079179$ & $-0.0852165$\\ \hline
$0.7$    & $-0.0601211$ & $-0.0642173$ & $-0.0687327$ & $-0.0737107$ & $-0.0791967$ & $-0.0852387$\\ \hline
$0.8$    & $-0.0599105$ & $-0.064092$ & $-0.0686764$ & $-0.0737036$ & $-0.0792162$ & $-0.0852609$\\ \hline
$0.9$    & $-0.0597946$ & $-0.0640191$ & $-0.0686449$ & $-0.0737046$ & $-0.0792353$ & $-0.0852802$\\ \hline
$0.99$   & $-0.0599869$ & $-0.0641012$ & $-0.0686779$ & $-0.0737213$ & $-0.0792489$ & $-0.0852906$\\ \hline
$0.999$  & $-0.060268$ & $-0.064232$ & $-0.0687298$ & $-0.0737374$ & $-0.0792521$ & $-0.085291$\\ \hline
$0.9999$ & $-0.060548$ & $-0.0643622$ & $-0.0687812$ & $-0.0737529$ & $-0.0792548$ & $-0.0852909$\\ \hline
\end{tabular}}

\vspace{0.2cm}

\centerline{\small\begin{tabular}{|c|c|c|c|c|c|c|} \hline
\multicolumn{7}{|c|}{$\Delta \Omega[\Phi_{\rm DW}(\overline{x}^1;\gamma)]$} \\ \hline
$\gamma$ & $\sigma=2.0$ & $\sigma=2.1$ & $\sigma=2.2$ & $\sigma=2.3$ & $\sigma=2.4$ & $\sigma=2.5$ \\ \hline
$0.01$   & $-0.0918881$ & $-0.0994172$ & $-0.107787$ & $-0.117054$ & $-0.127278$ & $-0.138519$\\ \hline
$0.1$    & $-0.0918881$ & $-0.0994119$ & $-0.107773$ & $-0.117029$ & $-0.127238$ & $-0.138463$\\ \hline
$0.2$    & $-0.0918881$ & $-0.0993961$ & $-0.107732$ & $-0.116954$ & $-0.127121$ & $-0.138294$\\ \hline
$0.3$    & $-0.0918881$ & $-0.0993705$ & $-0.107666$ & $-0.116833$ & $-0.126931$ & $-0.138021$\\ \hline
$0.4$    & $-0.0918881$ & $-0.0993362$ & $-0.107578$ & $-0.116672$ & $-0.126678$ & $-0.137659$\\ \hline
$0.5$    & $-0.0918881$ & $-0.0992947$ & $-0.107472$ & $-0.116478$ & $-0.126375$ & $-0.137226$\\ \hline
$0.6$    & $-0.0918881$ & $-0.0992481$ & $-0.107353$ & $-0.116263$ & $-0.126039$ & $-0.136747$\\ \hline
$0.7$    & $-0.0918881$ & $-0.0991994$ & $-0.107230$ & $-0.116040$ & $-0.125693$ & $-0.136255$\\ \hline
$0.8$    & $-0.0918881$ & $-0.0991525$ & $-0.107112$ & $-0.115828$ & $-0.125366$ & $-0.135794$\\ \hline
$0.9$    & $-0.0918881$ & $-0.0991132$ & $-0.107014$ & $-0.115655$ & $-0.125103$ & $-0.135431$\\ \hline
$0.99$   & $-0.0918881$ & $-0.0990933$ & $-0.106967$ & $-0.115580$ & $-0.125008$ & $-0.135338$\\ \hline
$0.999$  & $-0.0918881$ & $-0.0990931$ & $-0.106969$ & $-0.115592$ & $-0.125048$ & $-0.135436$\\ \hline
$0.9999$ & $-0.0918881$ & $-0.0990931$ & $-0.106972$ & $-0.115605$ & $-0.0125091$ & $-0.135538$\\ \hline
\end{tabular}}

\vspace{0.3cm}

\centerline{\small\begin{tabular}{|c|c|c|c|c|c|c|} \hline
\multicolumn{7}{|c|}{$\Delta \Omega[\mathbf{\Phi}_{\rm DW}(\overline{x};\gamma)]$} \\ \hline
$\gamma$ & $\sigma=2.6$ & $\sigma=2.7$ & $\sigma=2.8$ & $\sigma=2.9$ & $\sigma=3.0$ & $\sigma=3.1$ \\ \hline
$0.01$   & $-0.150841$ & $-0.164307$ & $-0.178984$ & $-0.194941$ & $-0.212247$ & $-0.230975$\\ \hline
$0.1$    & $-0.150764$ & $-0.164208$ & $-0.178860$ & $-0.194789$ & $-0.212065$ & $-0.230760$\\ \hline
$0.2$    & $-0.150536$ & $-0.163912$ & $-0.178489$ & $-0.194335$ & $-0.211522$ & $-0.230122$\\ \hline
$0.3$    & $-0.150168$ & $-0.163435$ & $-0.177892$ & $-0.193606$ & $-0.210650$ & $-0.229097$\\ \hline
$0.4$    & $-0.149678$ & $-0.162803$ & $-0.177100$ & $-0.192642$ & $-0.209499$ & $-0.227746$\\ \hline
$0.5$    & $-0.149094$ & $-0.162049$ & $-0.176160$ & $-0.191498$ & $-0.208138$ & $-0.226154$\\ \hline
$0.6$    & $-0.148451$ & $-0.161222$ & $-0.175131$ & $-0.190252$ & $-0.206659$ & $-0.224433$\\ \hline
$0.7$    & $-0.147794$ & $-0.160381$ & $-0.174091$ & $-0.188999$ & $-0.205185$ & $-0.222730$\\ \hline
$0.8$    & $-0.147183$ & $-0.159608$ & $-0.173147$ & $-0.187878$ & $-0.203887$ & $-0.221258$\\ \hline
$0.9$    & $-0.146715$ & $-0.159035$ & $-0.172474$ & $-0.187120$ & $-0.203064$ & $-0.220399$\\ \hline
$0.99$   & $-0.146660$ & $-0.159074$ & $-0.172683$ & $-0.187598$ & $-0.203935$ & $-0.221816$\\ \hline
$0.999$  & $-0.146863$ & $-0.159444$ & $-0.173306$ & $-0.188582$ & $-0.205415$ & $-0.223948$\\ \hline
$0.9999$ & $-0.147070$ & $-0.159821$ & $-0.173937$ & $-0.189575$ & $-0.206901$ & $-0.226091$\\ \hline
\end{tabular}}
\caption{\small One-loop $\mathbf{\Phi}_{\rm DW}(\overline{x}^1;\gamma)$-domain wall (non dimensional) tension shifts for several values of the family parameter $\gamma$ in the coupling constant range $\sigma\in [1.4,3.1]$.}
\end{table}

\begin{figure}[h]
\centerline{\includegraphics[height=3.5cm]{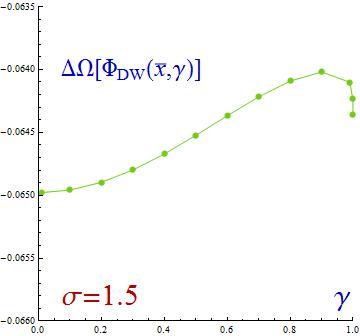} \hspace{0.5cm}
\includegraphics[height=3.5cm]{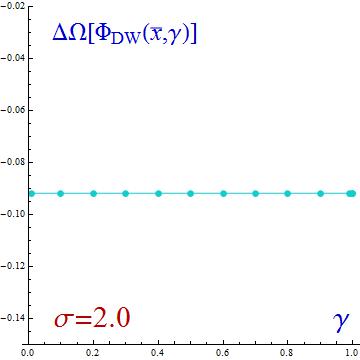} \hspace{0.5cm}  \includegraphics[height=3.5cm]{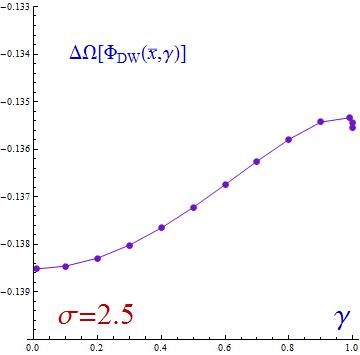}}
\caption{\small One-loop surface tension shifts for the cases, (a) $\sigma=1.5$, (b) $\sigma=2.0$, (c) $\sigma=2.5$ domain walls.}
\end{figure}

\section{Summary and future outlooks}

We have described a process of classical degeneracy breaking at the quantum level of the energy per unit of d-volume of the defects in a family of BPS-${\rm p}$-branes existing in a model of two real scalar fields in the particular cases of ${\rm p}=0$ -kinks- and ${\rm p}=2$ -domain walls. Each topological defect in the family is composed of two twin basic lumps separated by a certain distance. The classical degeneracy in the ${\rm p}$-brane tension amounts to the lack of interactions between the two constituent branes, independently of their separation. The quantum ${\rm p}$-brane fluctuations modify this situation because the one-loop ${\rm p}$-brane tension shifts differ with the relative position of the two basic branes, the parameter $\gamma$. In the $\sigma<2$ coupling constant regime, if ${\rm p}=0$, the two kinks repel each other, whereas if $\sigma>2$ the nature of the inter-kink forces depends on the distance: the force is attractive if the two lumps are close enough and it is repulsive otherwise. This bizarre behavior at large distances probably has to do with the fact that an infinite separation requires a change of topological sector in the configuration space. For the special case $\sigma=2$, the kink mass classical degeneracy survives, at least at one-loop order, in the quantum context. A phase transition at the critical value $\sigma=2$ converts the attractive force between two extended objects into a repulsive one. If ${\rm p}=2$ and we deal with BPS domain walls the results about one-loop shifts to the surface tension are qualitatively similar. There is also no saturation of the BPS bound at one-loop level, except if $\sigma=2$, although the quantum corrections to the wall surface tension are milder with respect to the kink mass shifts.

The forcefulness of these arguments to settle this picture is due to the precise computation of one-loop kink mass and domain wall tension shifts. The DHN formula is of no use in general because the spectral information available on the $2\times 2$-matrix differential operators governing the kink fluctuations is grossly insufficient. Thus, in previous publications alternative routes starting from the standard Gilkey-DeWitt heat kernel expansion were used. These methods are very  well adapted to dealing with ultraviolet divergences, but infrared phenomena are out of control. The existence of zero modes in both the kink and domain wall fluctuation spectra forbade a sufficiently precise response to answer the question about degeneracy breaking. In this paper we have modified the Gilkey-DeWitt heat trace expansion by taking into account the impact of zero modes at low temperatures adapted to field theory models with two scalar fields. This conceptual advance brought with it an improvement in the precision attained that allowed us to reach the results summarized in the previous paragraph. The extension of this procedure to compute one-loop surface tension shifts
 of classically degenerate domain walls -exhibiting also zero modes- benefitted from the previously mentioned conceptual advances in the treatment of kink fluctuations controlling the impact of zero modes.

The remarkable gain in precision achieved by building a heat kernel expansion that is also valid in the low temperature regime even when zero modes are present suggests that this method can be successfully applied to other problems in
one-loop physics. It is expected to work properly in other QFT systems such as gauge theories at zero and finite temperature. We mention a few prospects:

\vspace{0.2cm}

\noindent (1) The heat kernel expansion is also an effective tool in statistical physics. In Reference \cite{Megias2004} it is put to use to find the one-loop effective action in a kind of stochastic quantization of QCD. The authors did not consider Schr$\ddot{\rm o}$dinger operators with zero modes, but choosing their scalar field background as an extended object, e.g. a two-brane, the application of our modified GDW heat kernel expansion might be profitable.

\vspace{0.2cm}

\noindent (2) Semi-local self-dual topological solitons arising in the generalized Abelian Higgs model were studied in \cite{Alonso2008a}. The moduli space of these BPS solutions can be thought of as having a two-component boundary. In the first component  one finds all the Abrikosov-Nielsen-Olesen vortices \cite{Alonso2004,Alonso2005}. The second component is formed by the $\mathbb{CP}^1$-lumps, see \cite{Gibbons}. The self-dual semi-local topological solitons are hybrid objects that interpolate between the two extremes. In \cite{Alonso2008a} we used the standard GDW heat trace expansion to find that the one-loop fluctuations decreased the energy maximally for the pure ANO topological vortices. It is now tempting to rework the calculations relying on the modified heat kernel expansion in order to capture the low temperature effects.

\vspace{0.2cm}

\noindent (3) In all the physical problems mentioned up to here the important mathematical object is the spectral zeta function which is formally the $L^2$-trace of the complex power $-s$ of operators of Laplace, Dirac or Klein-Gordon type. Casimir energies, effective actions, etcetera, are thus regularized and, after proper renormalizations, evaluated. There are tunnel effect phenomena, the decay of false vacua \cite{Coleman}, instanton physics and the like where the solution of the conceptual conundrum requires the evaluation of functional determinants, which in turn are defined as the exponential of the derivative of the spectral zeta function at $s=0$. Because instantons and bounces have zero modes it is plausible that calculations of tunnel determinants, see e.g. \cite{Wipf1986}, will be more reliable using the modified GDW expansion.

\end{document}